# Finding the temperature window for atomic layer deposition of ruthenium metal via efficient phonon calculations

Alexandr Fonari[a,*], Simon D. Elliott[b], Casey N. Brock[a], Yan Li[a], Jacob Gavartin[c], Mathew D. Halls[d]

[a] Schrödinger Inc., New York, NY 10036, United States

[b] Schrödinger GmbH, 63163 Mannheim, Germany

[c] Schrödinger Technologies Ltd, United Kingdom

[d] Schrödinger Inc., San Diego, CA 92121, United States

[*] Corresponding author. E-mail: alexandr.fonari@schrodinger.com. Address: 1540 Broadway, Floor 24, New York, NY 10036, US

**Abstract**

We investigate the use of first principles thermodynamics based on periodic density functional theory (DFT) to examine the gas-surface chemistry of an oxidized ruthenium surface reacting with hydrogen gas. This reaction system features in the growth of ultrathin Ru films by atomic layer deposition (ALD). We reproduce and rationalize the experimental observation that ALD of the metal from $RuO_4$ and $H_2$ occurs only in a narrow temperature window above 100°C, and this validates the approach. Specifically, the temperature-dependent reaction free energies are computed for the competing potential reactions of the $H_2$ reagent, and show that surface oxide is reduced to water, which is predicted to desorb thermally above 113°C, exposing bare Ru that can further react to surface hydride, and hence deposit Ru metal. The saturating coverages give a predicted growth rate of 0.7 Å/cycle of Ru. At lower temperatures, free energies indicate that water is retained at the surface and reacts with the $RuO_4$ precursor to form an oxide film, also in agreement with experiment. The temperature dependence is obtained with the required accuracy by computing Gibbs free energy corrections from phonon calculations within the harmonic approximation. Surface phonons are computed rapidly and efficiently by parallelization on a cloud architecture within the Schrödinger Materials Science Suite. We also show that rotational and translational entropy of gases dominate the free energies, permitting an alternative approach without phonon calculations, which would be suitable for rapid pre-screening of gas-surface chemistries.

## 1. Introduction

How a solid surface is chemically modified by exposure to a gaseous reagent is an important issue in surface science, heterogeneous catalysis, materials processing and high-tech manufacturing. The exposure-induced change in surface chemistry may be advantageous or deleterious. It may be the first step in the reaction of the gas with the bulk of the solid material, or it may passivate the substrate with respect to this gas. In all cases, the main questions are what chemical intermediates are formed at the surface and to what extent the reaction can be controlled by the conditions of exposure.

An important example is atomic layer deposition (ALD), a technique for fabricating uniform thin solid films from gaseous reagents ('precursors') with unparalleled control of composition, thickness and conformality at the nanoscale.[1] This requires the growing surface to show self-limiting reactivity with respect to each precursor when pulsed separately, producing a surface-limited amount of deposited material in each cycle of precursor pulses. This distinguishes ALD from the wider family of chemical vapor deposition (CVD) processes, where growth is continuous with precursor exposure. In a similar way, the cyclic volatilization and passivation of surface intermediates by pulses of reagent gases facilitates atomic layer etch, whereas non-passivating etchant gases result in continuous etching. Similar questions about how gases transform surfaces arise for other processes, such as heterogeneous catalysis.

Aaltonen *et al.* pioneered the ALD of noble metal films by adsorbing reducing agent precursors such as $RuCp_2$ on surfaces oxidized by an O-based co-reagen,[2] generally requiring temperatures of at least 300°C to obtain Ru metal.[1] By contrast, Gatineau *et al.* introduced an oxidizing precursor $RuO_4$, along with $H_2$ as reducing agent, and thus lowered the deposition threshold to 200°C or less.[3] Pulsing $RuO_4$ at 0.0045 mbar and $H_2$ at 4 mbar, Minjauw *et al.* obtained genuine thermal ALD at 1.0 Å/cycle within a narrow temperature window around 100°C.[4] By observing that the thermal process at 75°C yielded O-rich films while $H_2$-plasma produced Ru metal films at this temperature, Minjauw *et al.* showed that the temperature window was the result of chemistry during the $H_2$ pulse.[5] Above the ALD temperature window, the decomposition of $RuO_4$ allows single-source deposition of oxide, or CVD of metal when $H_2$ is co-flowed.

Minjauw *et al.* propose the following mechanism, which is illustrated in Figure 1. In the metal precursor pulse, chemisorption of $RuO_4$ to metallic Ru (or to other oxidizable substrates) self-limits once the surface is saturated with oxide such as $RuO_2$. During the $H_2$ pulse, the oxide layer is then removed as water, restoring oxidizable Ru. The authors suggest that at 75°C the *kinetics* of reduction by $H_2$ are too slow for oxygen to be entirely removed, but that enough metal is formed to allow chemisorption of $RuO_4$ in the next ALD cycle. In this paper, we investigate whether the temperature window for Ru ALD can be explained instead by the *thermodynamics* of competing modes of reduction of a ruthenium oxide surface by $H_2$, rather than by the temperature-dependent kinetics of a single reaction.

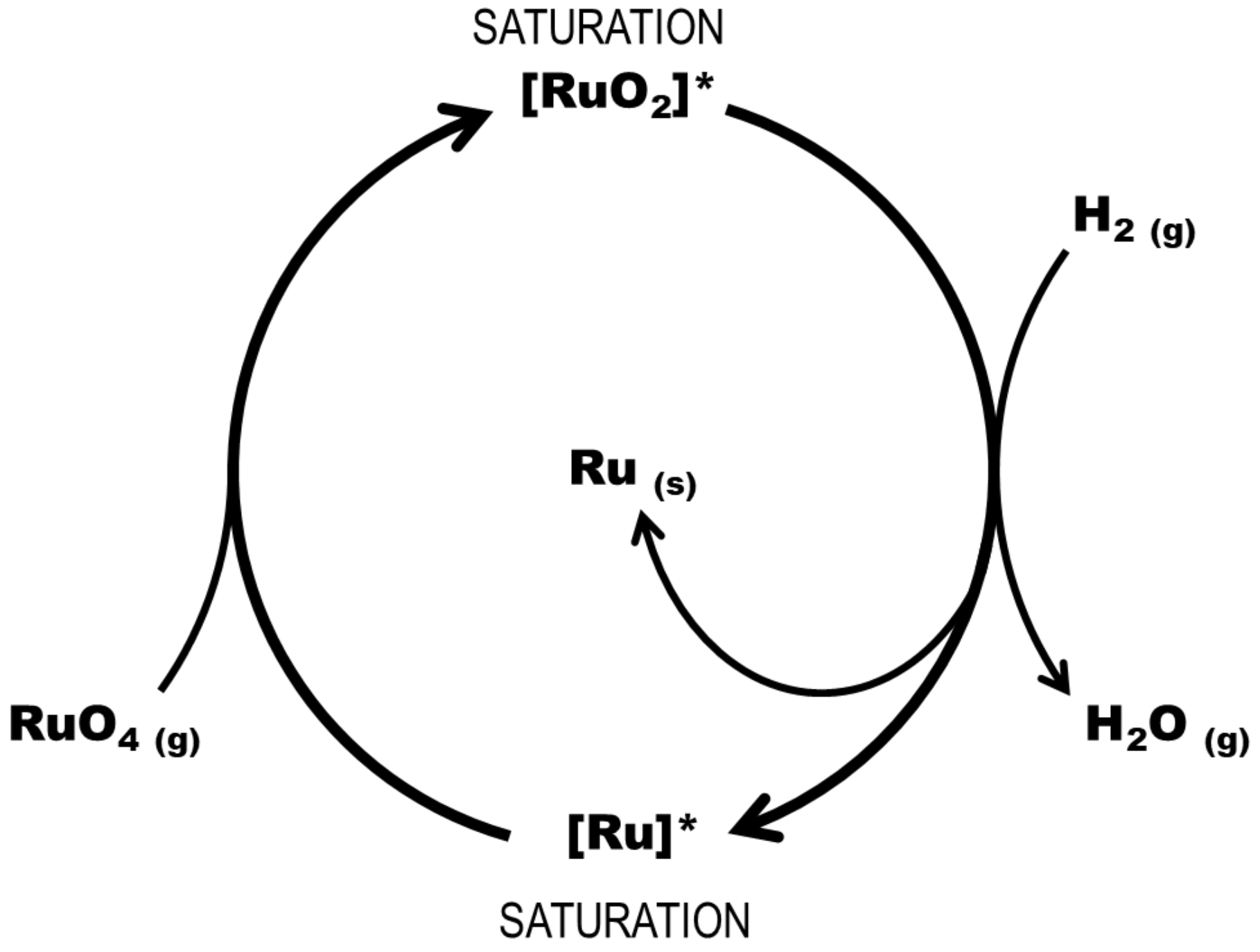

**Figure 1.** Schematic of mechanism proposed by Minjauw *et al.* [4] for thermal ALD of Ru metal at 100°C. At lower temperatures it is suggested that slow kinetics of reduction by $H_2$ is responsible for the observed deposition of $RuO_2$ instead of Ru. Gas species are denoted (g), bulk solids are denoted (s) and surface adsorbates are labeled with *.

In this study we use first principles thermodynamics [6] based on periodic density functional theory (DFT) to compute Gibbs reaction free energies and thus determine which gas-surface reactions dominate at particular temperatures. We look at various ways of computing the vibrational corrections that are critical for predicting Gibbs free energies. These thermal corrections are typically obtained for small systems using phonon calculations. However, phonon calculations are

extremely—often prohibitively—computationally expensive for the slab models of surfaces that are used with plane wave DFT. To address this challenge, we have implemented a cloud-enabled workflow, where phonon mode calculations are parallelized over many compute nodes. This allows turnaround of results in reasonable wall times and at lower financial cost through the use of pre-emptible compute nodes. We also assess and discuss a quicker but more approximate route to Gibbs free energies by treating gas molecules and surfaces as rigid.

## 2. Methods

### 2.1 Reaction free energy of gas adsorption on a surface

The reaction of a gas molecule *B* with a solid substrate *A* to give solid *C* and gaseous by-product *D* may be written as:

$$A_{(\mathrm{s})} + B_{(\mathrm{g})} \rightarrow C_{(\mathrm{s})} + D_{(\mathrm{g})} \qquad (1)$$

Eq. 1 is in general not an elementary step. In a harmonic approximation, the Gibbs reaction free energy ($\Delta_{\mathrm{r}}G$) is defined as:

$$\Delta_{\mathrm{r}}G(T) = \sum(E + ZPE + G_{\mathrm{corr}})_{\mathrm{products}} - \sum(E + ZPE + G_{\mathrm{corr}})_{\mathrm{reactants}}\,, \qquad (2)$$

where the total electronic energy (*E*) of the optimized chemical species, regardless of its type, solid or gas, is obtained directly from periodic DFT calculations. Zero point energy (*ZPE*) is defined as:

$$ZPE = \sum_i \frac{h\nu_i}{2}\,, \qquad (3)$$

where $\nu_i$ are the phonon frequencies.

Depending on the species type, the Gibbs free energy correction of gas ($G^{\mathrm{g}}_{\mathrm{corr}}$) and surface ($G^{\mathrm{s}}_{\mathrm{corr}}$) can be defined as a sum of several components: vibrational, rotational, translational, electronic, and configurational:

$$G^{\mathrm{g}}_{\mathrm{corr}} = G_{\mathrm{vib}} + G_{\mathrm{rot}} + G_{\mathrm{trans}} + G_{\mathrm{el}} + G_{\mathrm{conf}} \qquad (4a)$$

$$G^{\mathrm{s}}_{\mathrm{corr}} = G_{\mathrm{vib}} + G_{\mathrm{el}} + G_{\mathrm{conf}} \qquad (4b)$$

Note that rotational and translational components are assumed to be only present for isolated gas species.

Given the vibration frequencies $\nu_{\mathrm{i}}$, the vibrational component of the Gibbs free energy correction is:

$$G_{\mathrm{vib}} = k_{\mathrm{B}}T \sum_i \ln\left(1 - e^{-h\nu_i/k_{\mathrm{B}}T}\right), \qquad (5)$$

where $k_{\mathrm{B}}$ is Boltzmann constant, $h$ is Planck constant and $T$ is temperature. The rotational correction is:

$$G_{\mathrm{rot}} = -k_{\mathrm{B}}T \ln Q_{\mathrm{rot}}\ , \qquad (6)$$

where $Q_{\mathrm{rot}}$ is a rotational partition function that depends on the moment of inertia and symmetry number of the molecule.[7] The translational component is:

$$G_{\mathrm{trans}} = -k_{\mathrm{B}}T \ln\left[\left(\frac{2\pi M k_{\mathrm{B}}T}{h^2}\right)^{3/2} \frac{k_{\mathrm{B}}T}{P_0}\right] + k_{\mathrm{B}}T \ln\frac{P}{P_0}\ , \qquad (7)$$

where $M$ is molecular mass, $P$ is pressure and $P_0$ is reference pressure. As discussed in section 2.2, standard pressure is used throughout this paper, $P = P_0 = 1$ atm, so that the last term of Eq. 7 vanishes. The electronic correction is:

$$G_{\mathrm{el}} = -k_{\mathrm{B}}T \ln m\ , \qquad (8)$$

where $m$ is spin multiplicity. In this work only spinless systems are considered ($m = 1$) so that $G_{\mathrm{el}} = 0$ for all species, with the exception of the $O_2$ molecule (see supplementary information, SI), where the ground state triplet was calculated. Finally, we note that the configurational component of the free energy correction is not taken into account in this work, thus $G_{\mathrm{conf}} = 0$.

From the above equations and using data from the ground state phonon periodic calculations, it is possible to compute the Gibbs free energy of the reaction ($\Delta_{\mathrm{r}}G$) at given temperatures.

## 2.2 Selection of parameters to describe process conditions

The actual course of the process is determined by the kinetics of competing reactions, aggregated over the individual elementary steps, each under distinct conditions. If all elementary steps are known, techniques like microkinetic modelling can be used to solve for the process outcome (*i.e.* the pressure or coverage of all species under the specified conditions), but determining the elementary steps and computing the detailed pathway and activation energy for each one is a substantial task. The current approach aims to circumvent this by considering only aggregate reactions (not elementary steps) and how the underlying thermodynamics affect their competition, which is strictly valid only at equilibrium and after infinite time.

It is generally a good approximation to consider all species (gases and surfaces) to be at thermal equilibrium at the reactor temperature, although this clearly fails to describe cases where heat flow is significant.

The partial pressure of gases is more problematic to define. Isobaric ensembles operate under a constant total pressure. In a typical ALD reactor, a pulse of reactant gas is admitted within a flow of inert carrier gas, and this may be approximated as a partial pressure of reactant that is constant over the pulse duration and establishes an equilibrium with the surface. It is therefore possible to use thermodynamics to analyze the gas-surface equilibrium for particular partial pressures of reactants.[6] However the partial pressure of a product gas is usually not known and is probably not constant. A flow-type ALD reactor is an open system, where the product gas is evacuated through the flow of inert gas, potentially before equilibrium can be reached. The thermodynamic treatment that we present here is not therefore suitable for examining the pressure-dependence of reactions that feature such a product gas. Instead we omit the $k_B T \ln {}^{P}/_{P_0}$ term in Eq. 8 and restrict ourselves to standard state, $P = 1$ atm for all gas species.

### 2.3 Electronic structure method

All DFT calculations were performed using Quantum ESPRESSO version 7.3.1.[8] The Perdew Burke Ernzerhof (PBE) [9] functional was used to describe the electron exchange and correlation energies within the generalized gradient approximation for all systems. The D3 dispersion correction [10,11] was used to account for the van der Waals interactions. Computational details for bulk solids and gas molecules are given in section SI-1 of the Supplementary Information. The Brillouin zone was sampled using a 5×5×1 $k$-point mesh for all 2×2 surface cells. The ONCV norm conserving pseudopotentials set was used for the phonon calculations [12,13] with an energy cutoff of 60 Ry. When optimizing the structure, all atomic nuclei in the cell were relaxed until all atomic forces were below $1\times10^{-4}$ Ry/Bohr. The lower 5 layers of the slab were then kept fixed in the phonon calculations. Gas molecules were placed in a big enough box to prevent self interactions. Lattice vibrations (phonons) were computed using density functional perturbation theory (DFPT) [14] at the Γ-point.

## 3. Results and discussion

### 3.1 Intermediates and reactions

As outlined in section 1, there is strong experimental evidence that the narrow temperature window for thermal ALD is due to chemical reactions during the $H_2$ pulse. Nevertheless, we begin by checking the relative energetics of potential reactions of bulk Ru or $RuO_2$, including oxidation by $O_2$, $RuO_4$ or $H_2O$, reduction by $H_2$ or decomposition without a co-reagent (see supplementary information, section SI-1), and the results are consistent with the experimental finding. Reactions that show highly unfavorable thermodynamics for the bulk are not subsequently considered as candidates for surface reactions.

The next step is to generate structural models for surfaces covered with the proposed intermediates. The starting point is a fully-relaxed bare six-layer slab of (0 0 1)-oriented Ru, also denoted (0 0 0 1). The slab thickness in conjunction with a 10 Å vacuum gap ensures surface energy convergence to <0.1 $Jm^{-2}$. This bare surface is labeled *. A 2×2 expansion of the surface plane (5.4 Å × 5.4 Å) is used to investigate the coverage of a variety of possible surface intermediates during the ALD process, depicted in Figure 2.

Determining the optimum coverage of oxygen on the surface after oxidation in the $RuO_4$ ALD pulse is straightforward (section SI-2) and we find that energy is minimized for a Ru oxide adlayer with three oxygen atoms per 2×2 $Ru_4$ cell, denoted [3O]*. Reduction during the $H_2$ pulse can lead to a variety of intermediates. DFT-based energies were again minimized to find optimum coverages of hydride (four H per 2×2 cell, [4H]*) and of water (3 associated molecules per 2×2 cell, [3$H_2O$]*). Derived from these, we investigated a range of other potential Ru-O-H intermediates with DFT. We write down the following potential reactions by which the $H_2$ co-reagent can partially or completely reduce the [3O*] surface and generate the candidate intermediates during the $H_2$ ALD pulse.

**R1**: [3O]* + 3/2$H_2$ (g) → [3OH]*
**R2**: [3O]* + 3$H_2$ (g) → [3$H_2O$]*
**R3**: [3O]* + 3$H_2$ (g) → [$H_2O$]* + 2$H_2O$ (g)
**R4**: [3O]*+ 3$H_2$ (g) → * + 3$H_2O$ (g)
**R5**: [3O]* + 5$H_2$ (g) → [4H]* + 3$H_2O$ (g)

Many of these Ru-O-H intermediates are close in energy, so that an accurate calculation of free energy is needed for determining which surface predominates under experimental conditions. In section 3.3 reaction free energies for the above reactions are computed and their dependence on temperature is discussed, so as to determine the intermediates and reaction mechanism during the $H_2$ pulse of the ALD process.

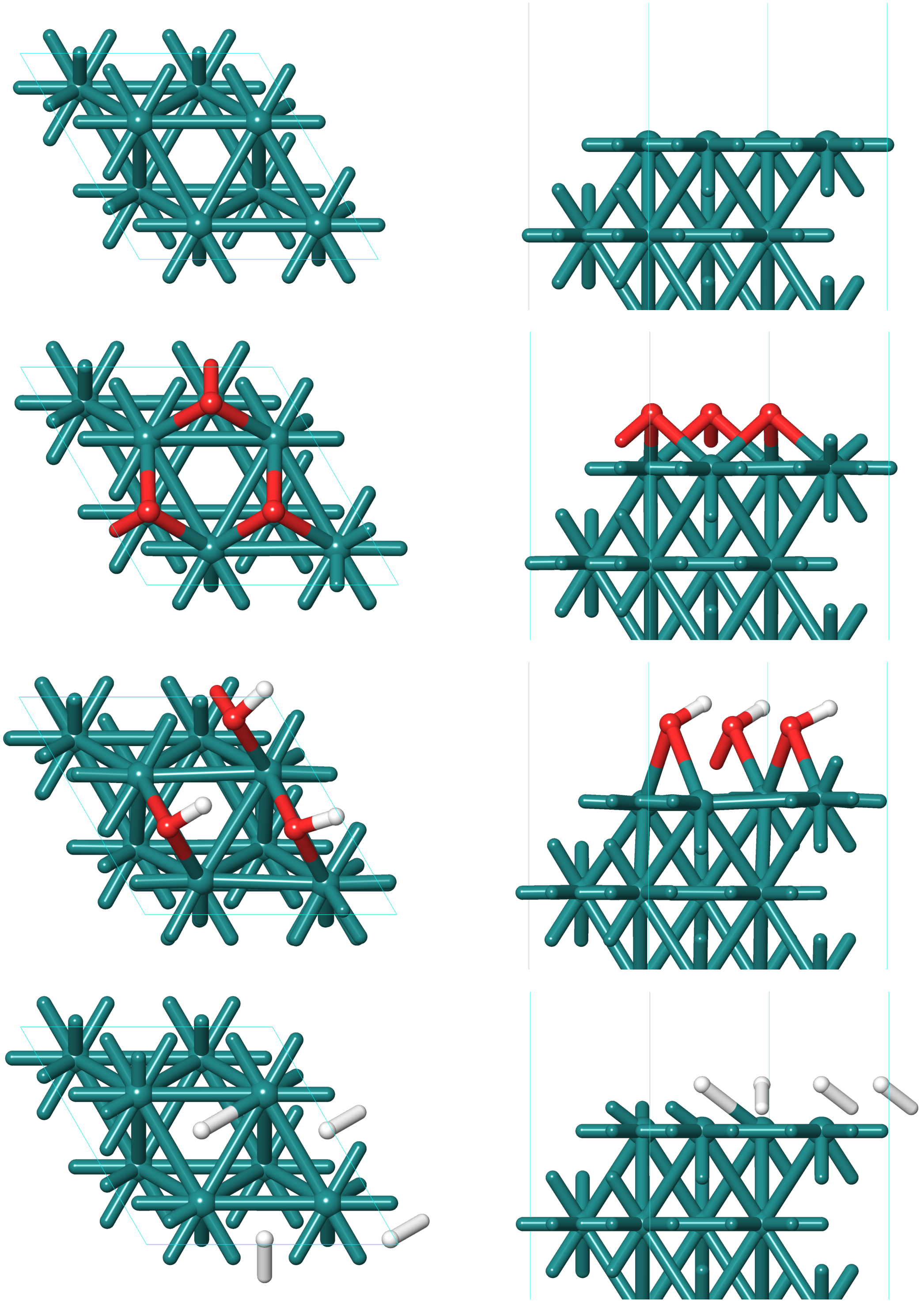

(a)

(b)

(c)

(d)

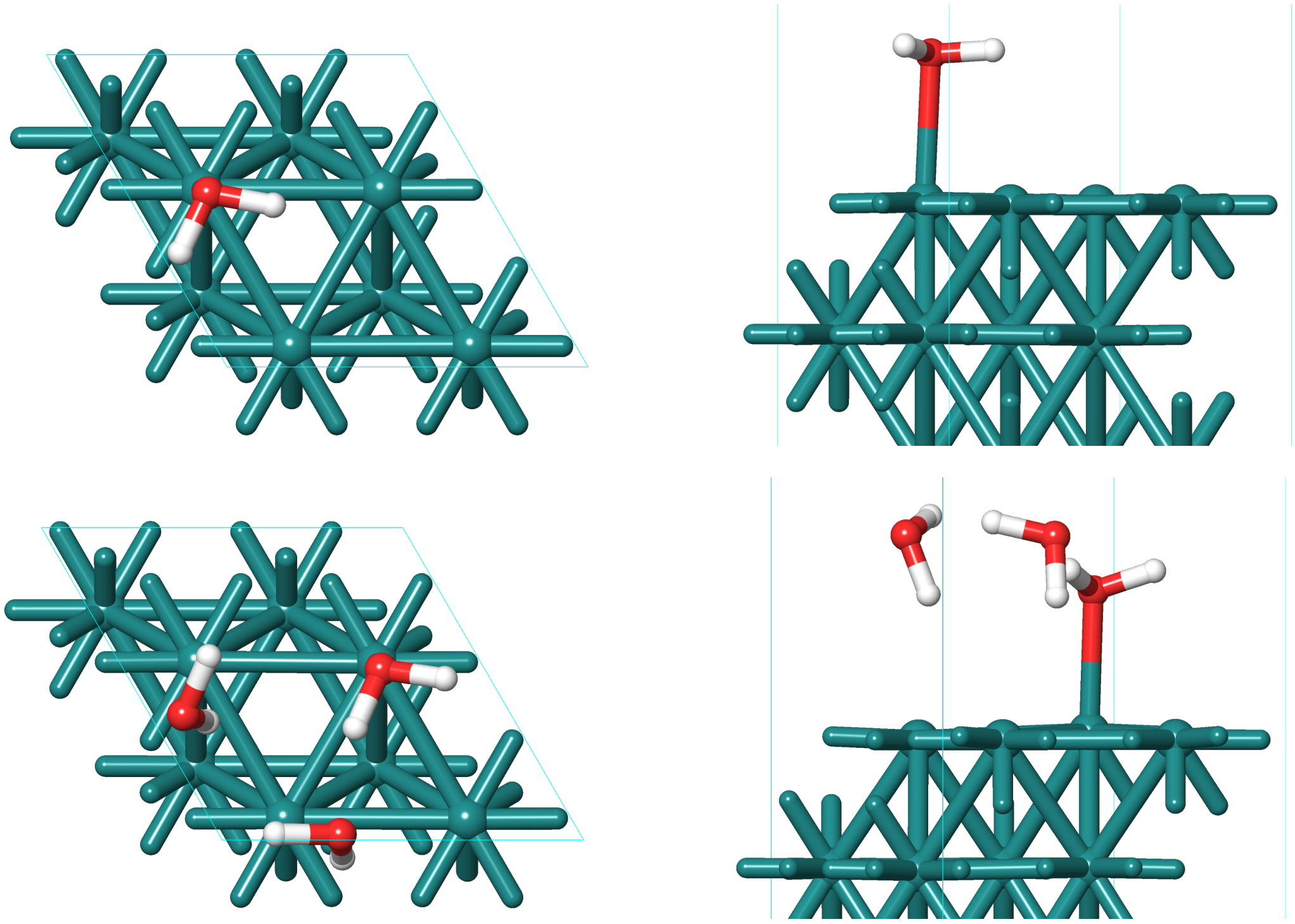

(e)

(f)

**Figure 2.** Top and side views of single unit cell of DFT-optimized slab structures of saturating surfaces and intermediates: (a) * = $Ru_{24}$/cell, (b) [3O]* = $O_3Ru_{24}$/cell, (c) [3OH]* = $H_3O_3Ru_{24}$/cell, (d) [4H]* = $H_4Ru_{24}$/cell, (e) [$H_2O$]* = $H_2ORu_{24}$/cell, (f) [$3H_2O$]* = $H_6O_3Ru_{24}$/cell. Ru=turquoise, O=red, H=white. Periodic expansion views are in Figure SI-3 of Supplementary Information.

### 3.2 Timing data for calculations using the distributed phonons workflow

In this section we describe the distributed phonons workflow and report timing data for the phonon calculations. In the following section, the thermochemical values resulting from these calculations will be discussed.

For a surface model with the bottom layers fixed and possibly deposited species, phonon modes are classified with the irreducible representation of the lowest symmetry point group $C_1$. Thus, lattice distortions are generated for every unconstrained atom (available degrees of freedom). These distortions are independent, allowing us to efficiently parallelize the entire phonon computation. We have implemented this workflow within the Schrödinger Materials Science Suite.[15] Using a job queuing system, as implemented in the Suite, it is possible to take advantage of either supercomputer or cloud resources. We have used maximum parallelization and have run each irreducible representation on a separate cloud node with 256 GB RAM and a total number of 64 CPU cores (AMD(R) EPYC(R) CPU @ 2.20GHz). As can be seen from Table 1, wall times of the distributed phonons workflow for the studied slab systems are all between 1 and 4 hours, the slowest system being [3$H_2O$]*. For a particular system, the distributed phonon computation wall time should be approximately equal to the time taken for the slowest convergence of any one representation. To confirm this, a sequential phonon computation was performed on the [4H]* surface on one node. The same phonons were confirmed and the wall time on one node (34 hours and 37 minutes) was in close agreement with the sum of wall times from the distributed workflow (32 hours and 58 minutes), thus arriving at near linear scaling between number of irreducible representations and compute nodes for the [4H]* surface.

**Table 1.** Number of unconstrained degrees of freedom for each system, number of irreducible representations (# irreps), average computation time per irreducible representation and wall times of the distributed phonons workflow for the studied systems.

| Slab system | Unconstrained atoms per cell | # irreps | Average time per irrep ± std, h:m | Wall time, h:m |
|---|---|---|---|---|
| * | $Ru_4$ | 12 | 1:29 ± 0:08 | 1:46 |
| [3O]* | $O_3Ru_4$ | 21 | 1:42 ± 0:09 | 2:01 |
| [4H]* | $H_4Ru_4$ | 24 | 1:22 ± 0:13 | 1:50 |
| [3OH]* | $H_3O_3Ru_4$ | 30 | 1:37 ± 0:16 | 2:12 |
| [$H_2O$]* | $H_2ORu_4$ | 21 | 1:50 ± 0:24 | 3:18 |
| [3$H_2O$]* | $H_6O_3Ru_4$ | 39 | 1:58 ± 0:29 | 3:47 |

### 3.3 Reaction energies from phonon calculations

The terms for the reaction Gibbs free energy from the phonon calculations are listed in Table 2. As discussed in Section 2.1, total energies (*E*) and *ZPE* corrections are temperature independent, whereas the Gibbs free energy correction does depend on the temperature for both surface and gas-phase species.

**Table 2.** Total energy, zero point energy and the Gibbs free energy correction at various temperatures obtained from phonon calculations.

| System | $E$, Ry | $ZPE$, kJ/mol | $G_{\text{corr}}$ at $T$, kJ/mol | | |
|---|---|---|---|---|---|
| | | | $T$=0°C | $T$=150°C | $T$=300°C |
| * | -4528.018 | 10.222 | -17.952 | -41.274 | -69.897 |
| [3O]* | -4623.906 | 39.868 | -17.148 | -43.172 | -77.531 |
| [4H]* | -4532.867 | 80.570 | -16.111 | -39.348 | -70.151 |
| [3OH]* | -4627.458 | 120.210 | -21.78 | -54.319 | -97.13 |
| [$H_2O$]* | -4562.361 | 71.296 | -22.322 | -51.99 | -89.004 |
| [3$H_2O$]* | -4631.064 | 199.253 | -31.481 | -74.051 | -128.328 |
| $H_2$ (g) | -2.332 | 25.839 | -26.978 | -47.183 | -68.970 |
| $H_2O$ (g) | -34.301 | 54.685 | -41.684 | -70.747 | -101.671 |

The data from Table 2 is applied to Eqs. 1-8 to give the Gibbs free energy of each of the reactions **R1**-**R5** across the temperature range of interest, and this is listed in Table 3.

**Table 3.** Gibbs reaction free energies ($\Delta_r G$) at various temperatures for reactions **R1**-**R5** from the computed data in Table 2. The reaction energies are plotted in Figure 3.

| Reaction | | $\Delta_r G$, kJ/mol | | |
|---|---|---|---|---|
| | | $T$=0°C | $T$=150°C | $T$=300°C |
| **R1** | [3O]* + 3/2$H_2$ (g) → [3OH]* | 1.8 | 7.8 | 13.8 |
| **R2** | [3O]* + 3$H_2$ (g) → [3$H_2O$]* | -15.8 | -4.7 | 6.6 |
| **R3** | [3O]* + 3$H_2$ (g) → [$H_2O$]* + 2$H_2O$ (g) | -5.9 | -6.2 | -6.0 |
| **R4** | [3O]*+ 3$H_2$ (g) → * + 3$H_2O$ (g) | -2.9 | -8.9 | -14.3 |
| **R5** | $[3O]^* + 5H_2$ (g) → $[4H]^*$ + $3H_2O$ (g) | -44.9 | -40.8 | -35.9 |

Motivated by the bulk calculations (section SI-1), we assume that surface reduction proceeds initially by (i) the dissociative adsorption of $H_2$ to give surface intermediates consisting of both O and H, followed by (ii) desorption of $H_2O$. The computed free energies shown in Figure 3 allow us to determine which of the mixed Ru-O-H intermediates is most thermodynamically favored across the temperatures of interest.

We can see that at low temperature, up to a crossover at 113°C, the fully hydrated intermediate $[3H_2O]^*$ is slightly more stable than the surfaces where water has desorbed, *i.e.* than the partially-hydrated $[H_2O]^*$ surface and the bare surface *, which are roughly isoenergetic. Surface Ru atoms are in oxidation state zero in all three surfaces, which helps rationalize their closeness in stability. The hydroxylated intermediate $[3OH]^*$ is much less stable. The fully-hydrated surface $[3H_2O]^*$ is likely to resist further adsorption of $H_2$, since it is the most H-rich intermediate identified by DFT calculations. Reaction **R2** therefore describes the surface transformation during the $H_2$ pulse at low temperature. Assuming that this fully-hydrated surface persists into the next $RuO_4$ pulse, it is plausible that water is displaced by adsorbed precursor, yielding bulk and surface oxide according to:

**R6**: $[3H_2O]^*$ + 3/2.$RuO_4$ (g) → $[3O]^*$ + 3/2.$RuO_2$ (s) + $3H_2O$ (g).

The deposition reaction per cell over the entire ALD cycle is thus obtained by adding reaction equations **R2** and **R6**, as follows, yielding ruthenium oxide as the material deposited below the crossover temperature:

**R7**: 3/2[$RuO_4$ (g) + $2H_2$ (g) → $RuO_2$ (s) + $2H_2O$ (g)].

This reaction is computed to be exoergic over the temperatures of interest (reaction **B13** of Table SI-2). This result agrees with the experimental finding of O-rich films with thermal $H_2$ at 75°C.[5] A schematic of the process is shown in Figure 4a.

Next, we discuss reactivities at temperatures above the 113°C crossover (Figure 3). We observe that desorption of water to produce the pure Ru surface is the most thermodynamically favored route to oxygen removal (reaction **R4**). Furthermore, the hydride-covered surface [4H]* is ultimately the most stable under any conditions where $H_2$ is present. This suggests a final reduction step (iii) where $H_2$ dissociates at the bare Ru surface into [4H]*, which is presumably only possible during a thermal $H_2$ pulse after the bare surface is exposed, *i.e.* above 113°C. (By contrast, plasma-$H_2$ would probably produce [4H]* at any temperature). Reaction **R5** thus describes the full $H_2$ reaction at temperatures above the crossover. The [4H]* surface is highly oxidizable, and thus can be expected to readily adsorb $RuO_4$ precursor in the next ALD pulse:

**R8**: $[4H]^* + 5/4.RuO_4\ (g) \rightarrow [3O]^* + 5/4.Ru\ (s) + 2H_2O\ (g)$.

Summing equations **R5** and **R8** gives the following overall reaction:

**R9**: $5/4[RuO_4\ (g) + 4H_2\ (g) \rightarrow Ru\ (s) + 4H_2O\ (g)]$.

Reaction **R9** represents the Ru deposition reaction per cell over the entire ALD cycle for temperatures above 113°C. This reaction is computed to be exoergic (reaction **B10** in Table SI-2). The overall cycle of reactions is illustrated in Figure 4b.[16]

Balancing reaction equations for these saturating coverages thus yields a predicted deposition rate of 5/4 Ru per $Ru_4$ cell, *i.e.* 5/16=0.3 monolayers of Ru per ALD cycle. Scaling this by the interlayer spacing (*c*/2=2.1 Å) or the cube root of the lattice volume (2.4 Å) per bulk Ru atom gives a thickness increment of 0.7 Å/cycle. This is a relatively high growth rate for metal ALD, consistent with a surface hydride that is sufficiently stable to contribute to oxygen removal (steps 2b & 1a of Fig. 3 in Ref. 16). The experimental growth per cycle is slightly higher again (1.0 Å [4]), probably indicating a contribution from CVD, which is observed to dominate experimentally already from 150°C. Alternatively, the discrepancy may indicate that larger slab expansions are required for more accurate saturating coverages.

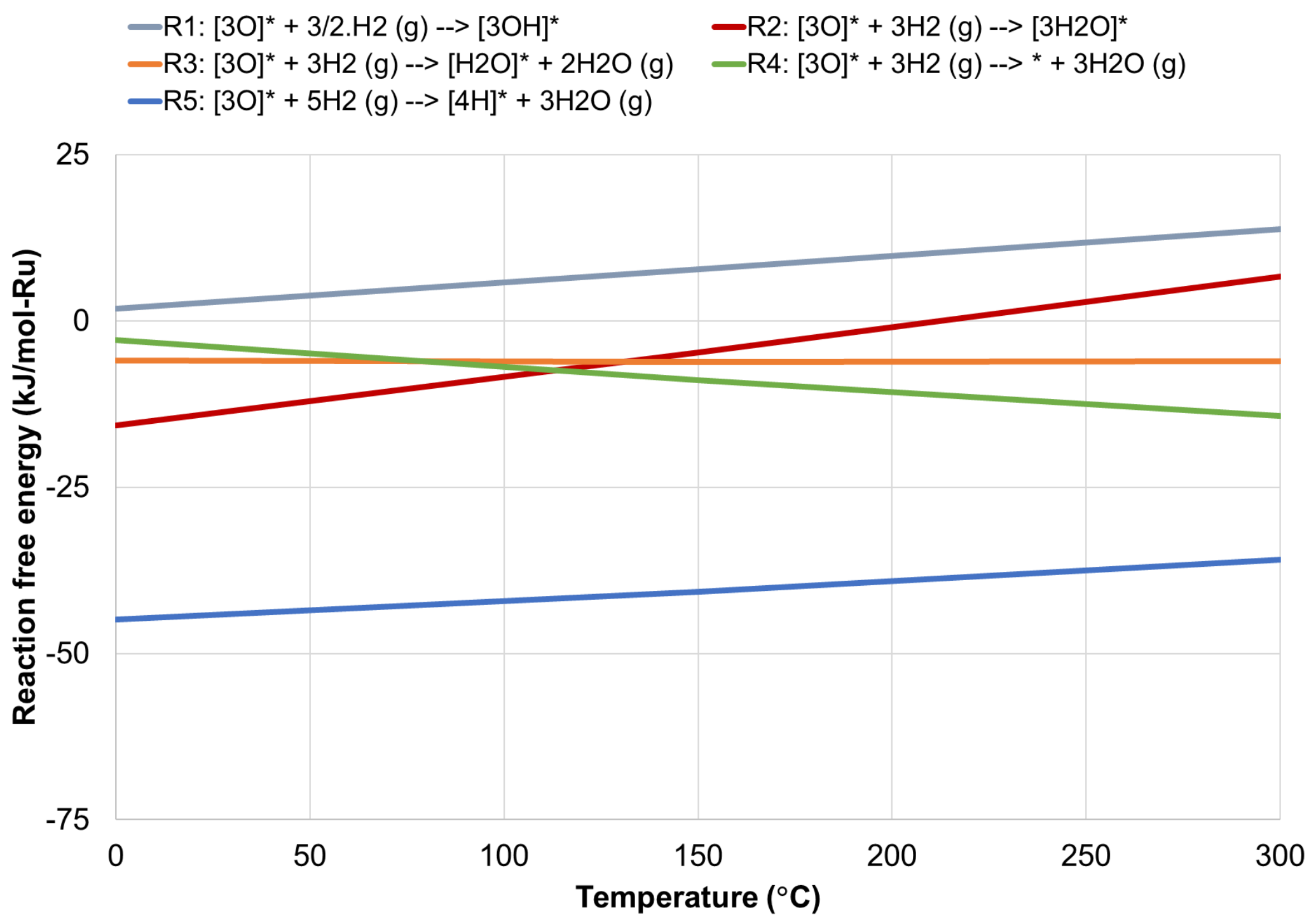
R1: [3O]* + 3/2.H2 (g) --> [3OH]*
R2: [3O]* + 3H2 (g) --> [3H2O]*
R3: [3O]* + 3H2 (g) --> [H2O]* + 2H2O (g)
R4: [3O]* + 3H2 (g) --> * + 3H2O (g)
R5: [3O]* + 5H2 (g) --> [4H]* + 3H2O (g)
Reaction free energy (kJ/mol-Ru)
25
0
-25
-50
-75
0
50
100
150
200
250
300
Temperature (°C)

(a)

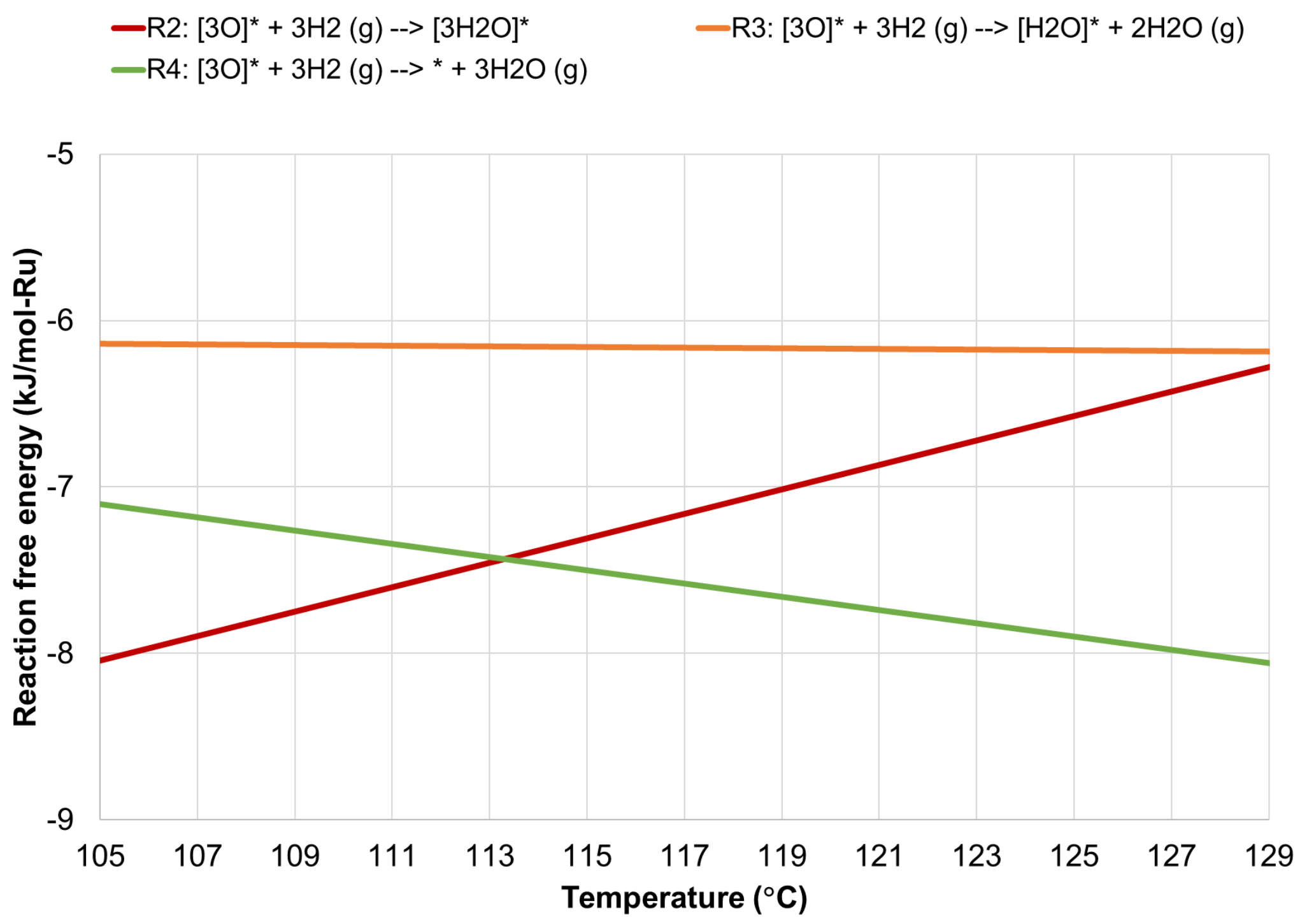
R2: [3O]* + 3H2 (g) --> [3H2O]*
R3: [3O]* + 3H2 (g) --> [H2O]* + 2H2O (g)
R4: [3O]* + 3H2 (g) --> * + 3H2O (g)
Reaction free energy (kJ/mol-Ru)
-5
-6
-7
-8
-9
105
107
109
111
113
115
117
119
121
123
125
127
129
Temperature (°C)

(b)

**Figure 3.** (a) Gibbs reaction free energies (Table 3) of competing reactions **R1**-**R5** of $H_2$ with oxidized Ru surface. (b) A magnified version that shows the crossover between **R2** and **R4** at 113°C.

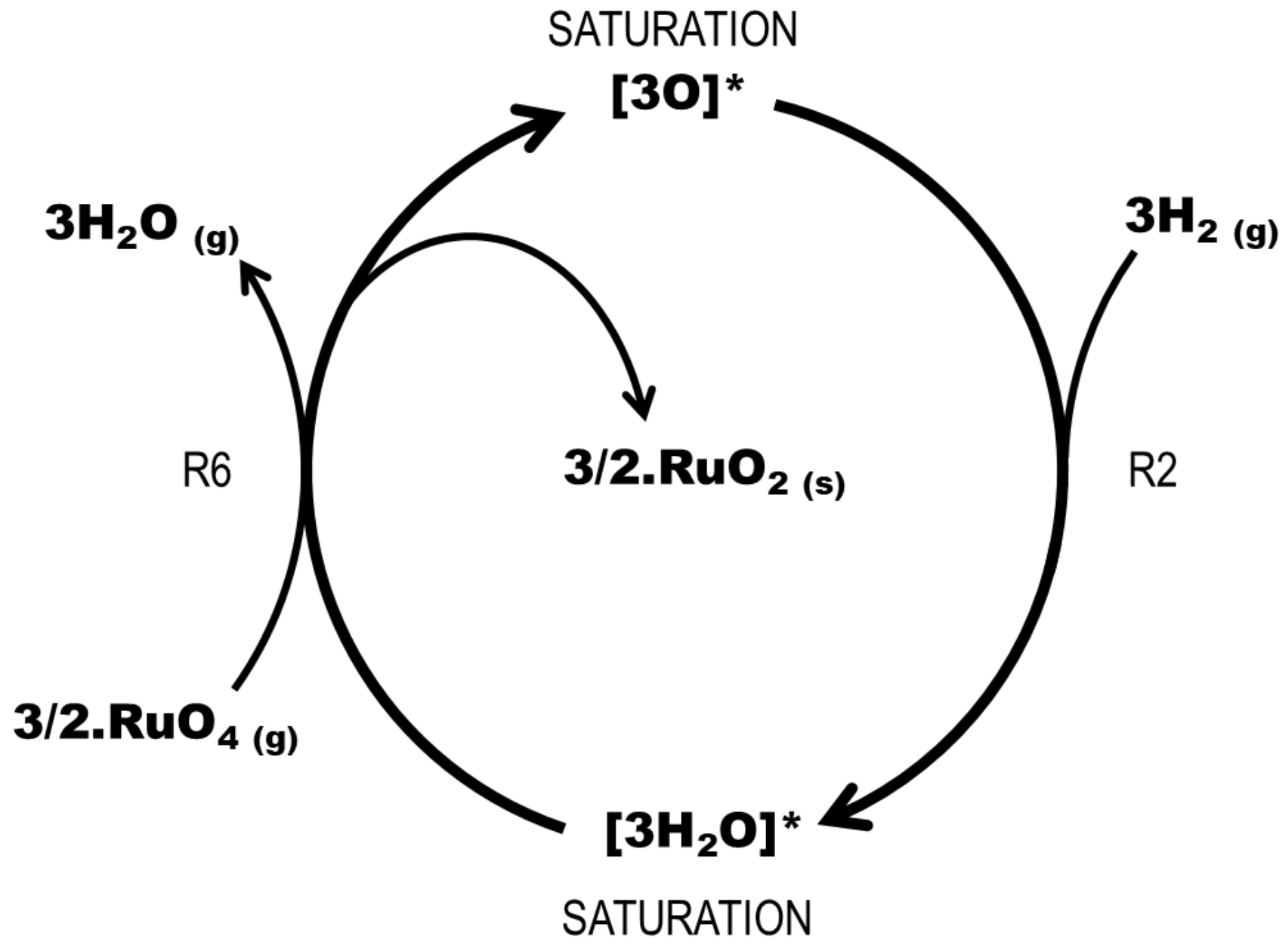

(a)

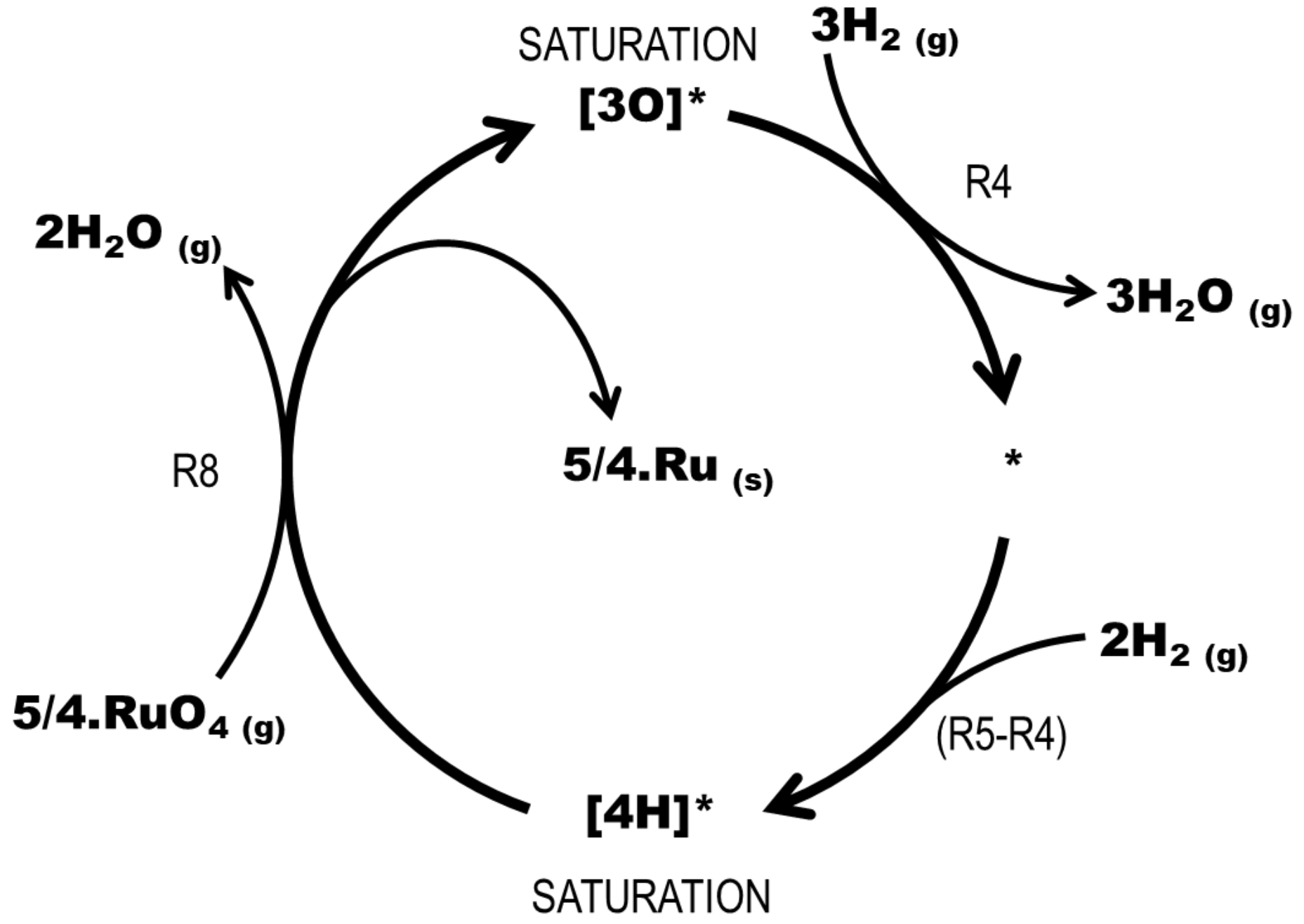

(b)

**Figure 4.** Mechanisms proposed here on the basis of thermodynamics of competing reactions

during the $H_2$ pulse (Figure 3): (a) deposition of ruthenium oxide below crossover temperature via retention of the hydrated surface at the end of the $H_2$ pulse; (b) ALD of ruthenium metal above crossover temperature via bare and then hydride-covered surface. Stoichiometric coefficients refer to quantities per (2×2) simulation cell with $Ru_4$ as the bare surface (*).

It was suggested in previous studies [4] that the temperature-dependence of the process was due to kinetic limitations that prevented the $H_2$ from converting the entire surface. By contrast, our study shows that thermodynamics alone can account for the observed temperature window, as a result of the different surface intermediates that occur above and below the crossover temperature.

### 3.4 Reaction energies using rigid approximation

It is possible to estimate Gibbs reaction energies without costly phonon calculations of the slabs, only using total electronic energies of the optimized species along with selected gas-phase contributions to free energy (primarily to entropy). The most straightforward approach is to set all the terms that depend on phonons to zero ($ZPE = G_{\mathrm{vib}} = 0$), which we denote as the 'rigid approximation'.[17-19] Thus from Eq. 2 one arrives to:

$$\Delta_{\mathrm{r}} G^{\mathrm{rigid}}(T) = \sum(E + G_{\mathrm{corr}}^{\mathrm{rigid}})_{\mathrm{products}} - \sum(E + G_{\mathrm{corr}}^{\mathrm{rigid}})_{\mathrm{reactants}} , \quad (9)$$

where rigid Gibbs free energy corrections for gas and surface system are defined respectively as:

$$G_{\mathrm{corr}}^{\mathrm{g,rigid}} = G_{\mathrm{rot}} + G_{\mathrm{trans}} \quad (10a)$$

$$G_{\mathrm{corr}}^{\mathrm{s,rigid}} = 0. \quad (10b)$$

Eq. 10b can occur when $\sum(G_{\mathrm{corr}}^{\mathrm{s,rigid}})_{\mathrm{products}} = \sum(G_{\mathrm{corr}}^{\mathrm{s,rigid}})_{\mathrm{reactants}}$ , *i.e.* by neglecting any changes in vibrational contribution to the Gibbs free energy of the surface as a result of adsorption or desorption reactions. $G_{\mathrm{rot}}^{\mathrm{g}}$ and $G_{\mathrm{trans}}^{\mathrm{g}}$ can be rapidly calculated from optimized geometries using a molecular or periodic DFT code.

Gibbs reaction free energies based on the rigid approximation are reported in Figure 5 and in the SI (Table SI-4) and show the crossover in surface reactivity at 108°C. This is remarkably close to the full phonon value (113°C). One reason for this is that the reactions have similar $\Delta_r ZPE$ values, within the range 16 ± 4 kJ/mol (Table SI-4). Setting $ZPE$ to zero in the rigid approximation thus produces a fairly uniform down-shift of all the reaction free energies compared to the case when phonons are taken into account. The second factor to consider is the temperature dependence of the free energy due to entropic contributions. The rigid approximation neglects all vibrational entropy and includes only rotational and translational gas-phase entropy. Analyzing the temperature-dependence of $\Delta_r G$ in Figure 3 and Figure 5 (see Table SI-4) confirms that this approximation quantitatively captures the trend in entropy, at least in this set of competing reactions and within the studied temperature range. We speculate that the rigid approximation would perform less well in the case of large and/or mobile adsorbates that substantially alter the vibrational entropy of the surface when they adsorb.

The rigid approximation is thus seen to capture the dominant contributions to the temperature-dependent free energy of these gas-surface reactions, while costing a fraction of the computational time of the phonon approach. This approximation may therefore be useful for rapid pre-screening of large numbers of candidate gas-surface reactions.

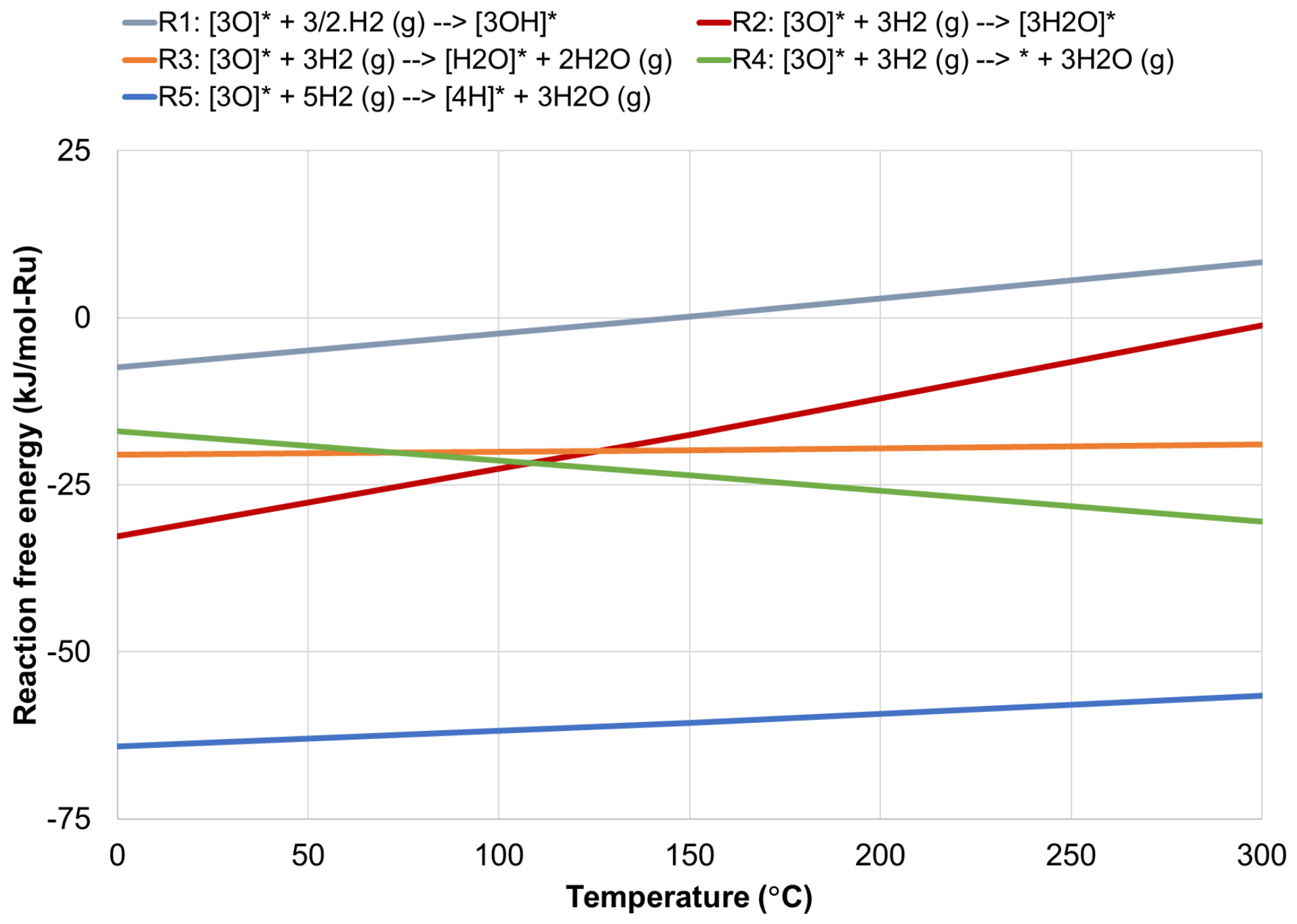
R1: [3O]* + 3/2.H2 (g) --> [3OH]*
R2: [3O]* + 3H2 (g) --> [3H2O]*
R3: [3O]* + 3H2 (g) --> [H2O]* + 2H2O (g)
R4: [3O]* + 3H2 (g) --> * + 3H2O (g)
R5: [3O]* + 5H2 (g) --> [4H]* + 3H2O (g)
Reaction free energy (kJ/mol-Ru)
Temperature (°C)

(a)

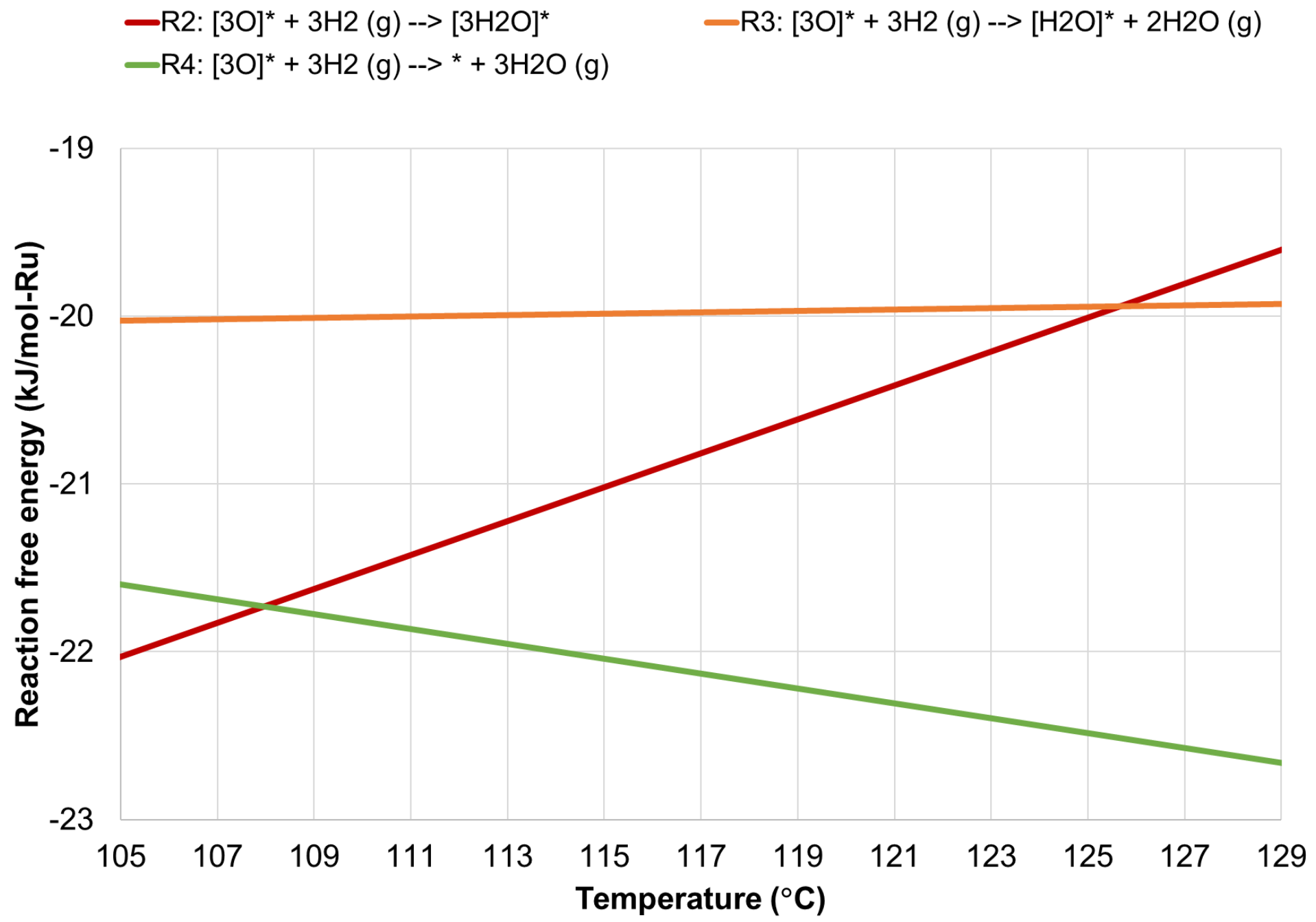
R2: [3O]* + 3H2 (g) --> [3H2O]*
R3: [3O]* + 3H2 (g) --> [H2O]* + 2H2O (g)
R4: [3O]* + 3H2 (g) --> * + 3H2O (g)
Reaction free energy (kJ/mol-Ru)
Temperature (°C)

(b)

**Figure 5.** Approximate Gibbs reaction free energies of competing reactions of $H_2$ with oxidized Ru surface within the 'rigid' approximation (see Eq. 9). (b) A magnified version that shows the crossover between **R2** and **R4** at 108°C.

## 4. Conclusions

In this paper we reproduced and rationalized the experimental observation that the atomic layer deposition (ALD) of ruthenium metal from $RuO_4$ and $H_2$ occurs only in a narrow temperature window above 100°C. Reduction during the $H_2$ pulse of ALD can lead to a variety of near iso-energetic Ru-O-H intermediates and distinguishing between them requires accurate calculation of the temperature-dependent free energy. The results show a crossover in free energies at 113°C for the onset of metal ALD, which we can thus ascribe to particular surface chemistry. Above the crossover temperature, thermal desorption of the water by-product from the surface is predicted, driving the equilibrium towards bare and then hydride-covered surfaces, which lead to the deposition of Ru metal. At temperatures below the crossover, water is retained at the surface after the $H_2$ pulse and reacts with the next pulse of $RuO_4$ to form an oxide film, also in agreement with experiment. The predicted crossover temperature of 113°C compares well with the experimental value of 100°C.

We note that kinetics usually plays a critical role in surface chemistry, especially when considering temperature-dependent reactivity. So, the activation energies of rate-limiting reaction steps would typically be required alongside reaction free energies for a full description. However, our calculations suggest that in this particular case of metal ALD, the experimentally observed temperature window for Ru deposition is purely the result of thermodynamics, that is, can be explained solely based on the reaction free energies, which significantly simplifies calculations.

We have established the importance and utility of accurately describing the temperature dependence of the free energy. Future work could consider how to compute pressure dependence in such reactions, which can be important in cases where by-products play a role. As discussed in section 2.2, the results presented here are for equal partial pressures of the two reagents and of the by-products, which could be an over-simplification of the situation in an ALD experiment. This limitation should be borne in mind when comparing to experiments at specific reagent partial pressures.

By quantifying the saturating coverages and associated reaction equations for the mechanism in each half-cycle, we can estimate the growth per cycle for the entire ALD cycle. Our prediction of 0.7 Å/cycle Ru metal above 113°C compares favorably with the experimental value of 1.0 Å/cycle at and above 100°C, which probably includes a CVD component. This level of agreement validates our DFT phonon-based free energy approach as an accurate and efficient way to analyze competing gas-surface chemistries across a wide range of technological applications.

Correctly describing reaction energetics is important for understanding and exploiting gas-surface chemistry, and is therefore indispensable for progress in areas such as heterogeneous catalysis, corrosion and thin film processing. Here we presented a workflow to compute gas-surface reaction thermodynamics by taking vibrations into account in periodic DFT calculations of phonons for both surfaces and gas molecules, implemented in a highly scalable cloud-friendly approach. We also assessed the accuracy of a rigid approximation that does not require phonon calculations.

**Conflicts of interest**

There are no conflicts to declare.

# Finding the temperature window for atomic layer deposition of ruthenium metal via efficient phonon calculations

Alexandr Fonari[a,*], Simon D. Elliott[b], Casey N. Brock[a], Yan Li[a], Jacob Gavartin[c], Mathew D. Halls[d]

[a] Schrödinger Inc., New York, NY 10036, United States

[b] Schrödinger GmbH, 63163 Mannheim, Germany

[c] Schrödinger Technologies Ltd, United Kingdom

[d] Schrödinger Inc., San Diego, CA 92121, United States

[*] Corresponding author. E-mail: alexandr.fonari@schrodinger.com. Address: 1540 Broadway, Floor 24, New York, NY 10036, US

## Supplementary Information

### Section SI-1: Workflow for bulk screening

The first step of the workflow was to compute electronic energies and free energies of bulk solids, reagent gases and by-product gases with the periodic code Quantum ESPRESSO (Table SI-1), so as to obtain reaction free energies ($\Delta G$) of a range of possible bulk-gas reactions under the conditions of interest (Table SI-2). Here, the free energies have been obtained from full phonon calculations, which are relatively quick for these small systems. Nevertheless, an alternative is to neglect bulk phonons and obtain approximate free energy differences from ‘rigid’ rotations and translations of gas-phase molecules only (section 3.4), which can be justified as long as the energy differences ($\Delta E$) between reactions are large. Either way, computing bulk solids and gases is an efficient way to screen out unlikely processes or mechanisms, reducing the scope of reactions that need to be studied in more detail at the surface.

The crystal structures of the bulk solids Ru (hcp $P6_3/mmc$) and $RuO_2$ (rutile $P4_2/mnm$) were optimized with the PBE-D3 functional,[1-3] GBRV ultrasoft pseudopotentials [4] and plane wave cutoffs of 40 Ry for wavefunctions and 200 Ry for charge density, with *k*-point grid planes spaced at 0.04/Å for Ru (10×10×6 Monkhorst-Pack mesh) and 0.10/Å for $RuO_2$ (6×6×8 mesh). The optimized lattice constants were $a$=$b$=2.701 Å, $c$=4.269 Å, α=β=90°, γ=120° for the $Ru_2$ cell and $a$=$b$=4.501 Å, $c$=3.131 Å, α=β=γ=90° for the $Ru_2O_4$ cell.

Molecular models of the reagent and by-product gases $H_2$, $H_2O$, $RuO_4$ and $O_2$ were generated within otherwise-empty cubic cells so that periodic images were at least 7 Å apart and the structures were optimized with the same functional, pseudopotentials and plane wave cutoffs as for the bulk, but at the Γ *k*-point only.

Phonon calculations (section 3.2) were performed in order to obtain the zero-point energy and free energy corrections for each bulk and gas-phase species (Table SI-1).

**Table SI-1.** Total energy, zero point energy and free energy correction from full phonon calculations at different temperatures of bulk and molecular systems.

| | ***E*, Ry** | ***ZPE*, kJ/mol** | **Free energy correction (kJ/mol) at *T* (K) and *P*=1 atm** | | |
|---|---|---|---|---|---|
| | | | ***T*=0°C** | ***T*=150°C** | ***T*=300°C** |
| **bulk-Ru, per Ru** | -188.700 | 1.958 | -1.313 | -3.437 | -6.205 |
| **bulk-$RuO_2$, per Ru** | -252.498 | 20.043 | -2.829 | -8.612 | -17.106 |
| **$H_{2\ (g)}$** | -2.332 | 25.839 | -26.978 | -47.183 | -68.97 |
| **$H_2O_{(g)}$** | -34.301 | 54.685 | -41.684 | -70.747 | -101.671 |
| **$RuO_{4\ (g)}$** | -316.052 | 32.869 | -63.488 | -108.376 | -157.839 |
| **$O_{2\ (g)}$ triplet** | -63.492 | 9.025 | -45.038 | -75.180 | -106.971 |

The thermodynamics of possible transformations of solid Ru are considered in Reactions **B1**-**B5**

of Table SI-2 and shown in Figure SI-1. The most favorable of these reactions involve oxidation by $O_2$ (**B1** & **B2**), but $O_2$ should be excluded from the ALD reactor when attempting to deposit Ru metal. Reaction **B3** is also exoergic, indicating that oxidation of bulk Ru metal by $RuO_4$ during precursor exposure is favorable, and most likely occurs at the surface as well. Bulk oxidation by the $H_2O$ by-product is computed to be endoergic (**B4** & **B5**) and so can be ignored as a potential reaction at the surface.

**Table SI-2.** Reaction free energies (kJ/mol-Ru) of possible reactions of the bulk solids Ru and $RuO_2$ and gas-phase precursor $RuO_4$ at temperature $T$ and $P$=1 atm using the full phonon free energy correction.

| Label | Reaction | $\Delta_r E$, kJ/mol | $\Delta_r G$, kJ/mol | | |
|---|---|---|---|---|---|
| | | | $T$=0°C | $T$=150°C | $T$=300°C |
| **B1** | $Ru_{(s)} + 2O_2 \rightarrow RuO_{4\,(g)}$ | -484.2 | -443.4 | -425.9 | -409.0 |
| **B2** | $Ru_{(s)} + O_{2\,(g)} \rightarrow RuO_{2\,(s)}$ | -401.5 | -348.9 | -322.4 | -296.4 |
| **B3** | $\frac{1}{2} Ru_{(s)} + \frac{1}{2} RuO_{4\,(g)} \rightarrow RuO_{2\,(s)}$ | -318.8 | -254.4 | -218.9 | -183.7 |
| **B4** | $Ru_{(s)} + 2H_2O_{(g)} \rightarrow RuO_{2\,(s)} + 2H_{2\,(g)}$ | 184.0 | 172.3 | 186.3 | 198.9 |
| **B5** | $Ru_{(s)} + 4H_2O_{(g)} \rightarrow RuO_{4\,(g)} + 4H_{2\,(g)}$ | 686.7 | 598.9 | 591.6 | 581.4 |
| **B6** | $RuO_{2\,(s)} + 2H_{2\,(g)} \rightarrow Ru_{(s)} + 2H_2O_{(g)}$ | -184.0 | -172.3 | -186.3 | -198.9 |
| **B7** | $RuO_{2\,(s)} + O_{2\,(g)} \rightarrow RuO_{4\,(g)}$ | -82.7 | -94.5 | -103.5 | -112.7 |
| **B8** | $RuO_{2\,(s)} \rightarrow Ru_{(s)} + RuO_{4\,(g)}$ | 318.8 | 254.4 | 218.9 | 183.7 |
| **B9** | $RuO_{2\,(s)} \rightarrow Ru_{(s)} + O_{2\,(g)}$ | 401.5 | 348.9 | 322.4 | 296.4 |
| **B10** | $RuO_{4\,(g)} + 4H_{2\,(g)} \rightarrow Ru_{(s)} + 4H_2O_{(g)}$ | -686.7 | -598.9 | -591.6 | -581.4 |
| **B11** | $RuO_{4\,(g)} \rightarrow RuO_{2\,(s)} + O_{2\,(g)}$ | 82.7 | 94.5 | 103.5 | 112.7 |
| **B12** | $RuO_{4\,(g)} \rightarrow Ru_{(s)} + 2O_{2\,(g)}$ | 484.2 | 443.4 | 425.9 | 409.0 |
| **B13** | $RuO_{4\,(g)} + 2H_{2\,(g)} \rightarrow RuO_{2\,(s)} + 2H_2O_{(g)}$ | -502.8 | -426.6 | -405.3 | -382.6 |

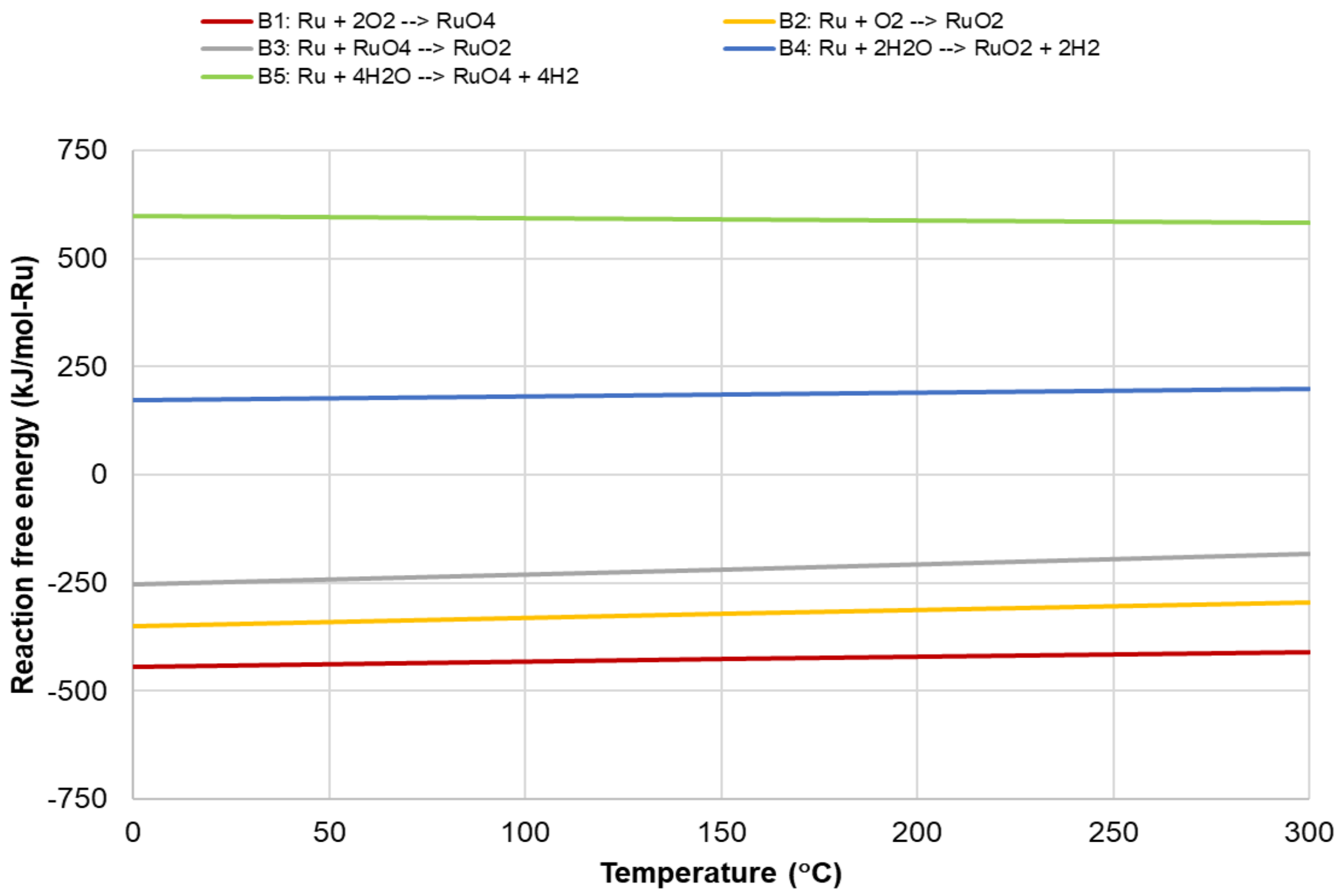

**Figure SI-1.** Computed free energies from Table SI-2 of potential oxidation reactions of bulk solid Ru with the gases that may be present, namely the precursor $RuO_4$ and the by-products $O_2$ and $H_2O$.

For bulk $RuO_2$, the negative free energy computed for reduction by $H_2$ to yield $H_2O$ (reaction **B6**) indicates that this could occur during the $H_2$ pulse of ALD, and this motivates our study of $H_2$ reactivity with oxidized Ru surfaces in this paper. The data for reactions **B8** and **B9** show that auto-reduction of $RuO_2$ in the absence of a reducing agent can not take place.

The reaction for deposition of Ru metal from precursor $RuO_4$ and co-reagent $H_2$ (**B10**) is computed to be highly exoergic across the entire temperature range. The competing reaction of oxide deposition (**B13**) is also exoergic. The overall reactions **B10** and **B13** are independent of mechanism, and therefore equally well describe CVD when co-flowing the reagents or ALD from alternating exposures. By contrast, single-source CVD of either $RuO_2$ or Ru through thermal decomposition of the precursor (reactions **B11** & **B12**) is seen to be endoergic, and can therefore be ignored.

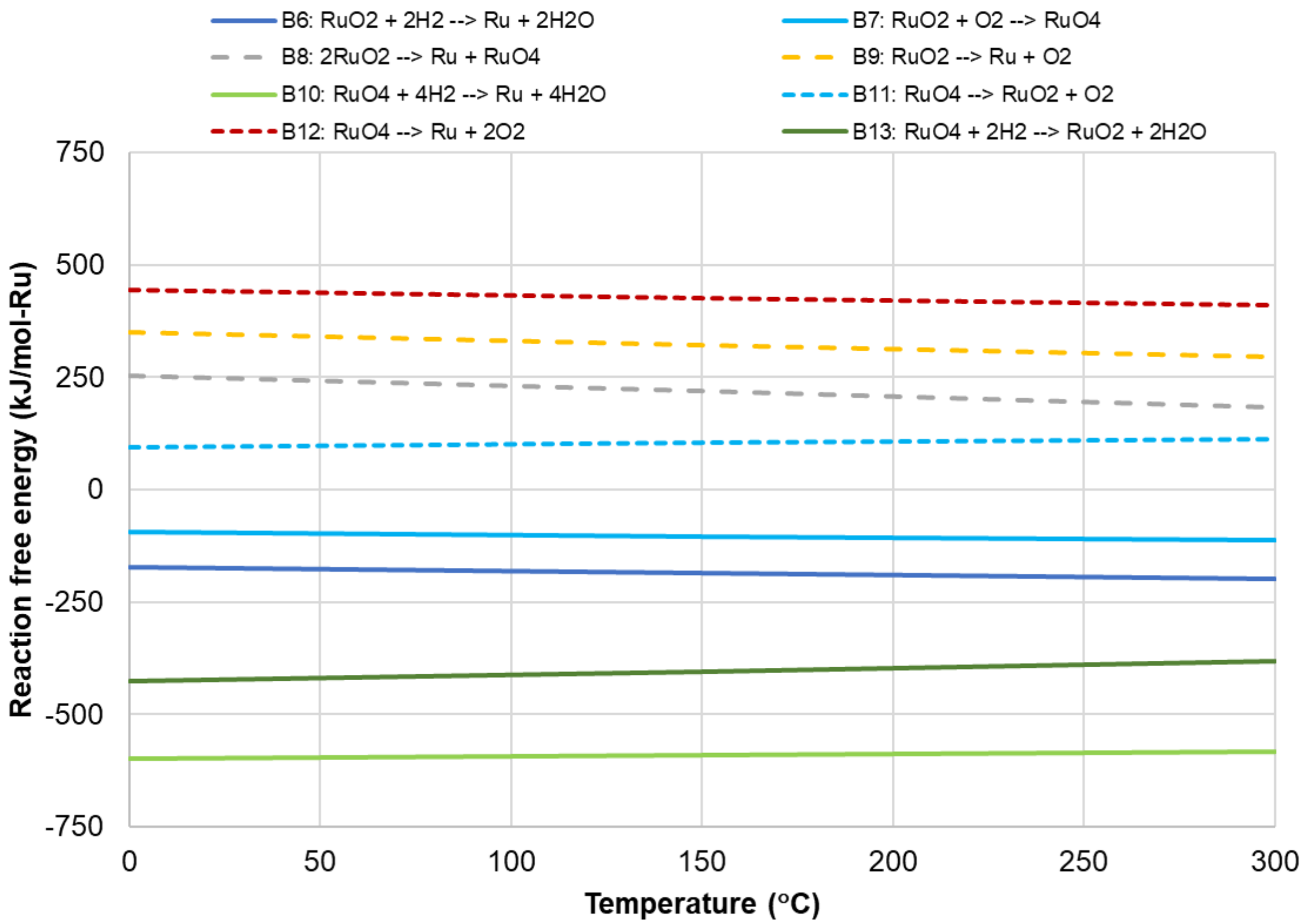

**Figure SI-2.** Computed free energies from Table SI-2 of potential reactions of bulk solid $RuO_2$ and of gaseous $RuO_4$, with dashed lines indicating potential auto-decomposition reactions.

## Section SI-2: Workflow for finding saturated surfaces

Saturated surfaces were generated by functionalizing the 2×2 expansion of a bare (0 0 1) Ru slab that is six $Ru_4$/cell layers thick and separated from images by 10 Å of vacuum. The bare surface is labeled as *.

In order to find a representative structural model of the oxidized surface after the $RuO_4$ ALD pulse, O atoms were added to one face of the bare slab and the geometry was relaxed with DFT. Coverages ranging 1(O) to 5(O) per $Ru_4$ cell and on-top, bridging and capping locations were investigated. On-top adsorption, sub-surface positions and peroxide dimers (O-O distance = 1.40 Å) were all found to be less favored energetically than locating O at a capping site, which relaxed to a position equidistant from three surface Ru atoms.

Table SI-3 summarizes the methods used to analyze the total energies of the optimized slabs so as to find the saturating surface. The results presented here are for $O_2$ gas as oxidant, but equivalent results can be obtained for the ALD precursor $RuO_4$ as oxidant. As shown in section SI-1, oxidation of bulk Ru by $O_2$ (or $RuO_4$) is energetically favorable and so it is not surprising that oxidation of the Ru surface by $O_2$ at $T$=0 K and $P$=1 atm is exoergic overall ($\Delta E_{avg}$<0) for all geometries, albeit decreasing per atom with coverage.  More useful is the energy for incremental oxidation by the $N^{th}$ O atom relative to the surface already oxidized by ($N$-1)O atoms. As shown in Table SI-3, this $\Delta E_N$ is negative for $N \leq 3$ and positive for $N \geq 4$, meaning that oxidation beyond 3(O) per $Ru_4$ cell is not energetically favorable. An alternative but equivalent approach is to find geometries where surface oxidation is favored relative to bulk oxidation. This is done by viewing the partially-oxidized slabs as mixtures of Ru metal and $RuO_2$ and computing their surface energies relative to bulk Ru and bulk $RuO_2$. Using this metric, Table SI-3 again shows that the most stable isomer at coverage of 3(O) per $Ru_4$ cell has the lowest surface energy, and is thus the saturating surface after oxidation. The structure is shown in Figure SI-3(a).  This can be rationalized as the maximum coverage that permits all adsorbed O to occupy hcp capping sites.

The change in $\Delta E$ from $N$=3 to $N$=4 is so substantial that it is not necessary to adjust for the free energy correction and consider $\Delta G$ as a function of temperature for the $RuO_4$ pulse. However, for the $H_2$ pulse, it is necessary to compute the temperature-dependent free energy in order to determine which Ru-O-H system is obtained, as various candidate surfaces lie close in energy (section 3 of main paper).

**Table SI-3.** Analysis of oxidized surfaces, where $N$ is the number of O atoms added per $Ru_4$ slab, $\Delta E_{avg}$ is the oxidation energy relative to $O_2$ gas averaged over all $N$(O), $\Delta E_N$ is the energy of oxidation by ½$O_2$ relative to the lowest energy slab with ($N$-1)O and $E_{surf}$ is the surface energy relative to $N$/2(bulk-$RuO_2$) and (24-$N$/2)(bulk-Ru).

| ***N*** | **Description of optimized geometry** | **$\Delta E_{avg}$ (kJ/mol-O)** | **$\Delta E_N$ (kJ/mol-O)** | **$E_{surf}$ (J/m$^2$)** |
|---|---|---|---|---|
| 1 | 1(hcp-cap) | -291.5 | -291.5 | 2.84 |
| 2 | 2(hcp-cap) | -257.6 | -233.6 | 2.60 |
| 2 | 1(fcc-cap)+1(sub-layer) | -122.5 | +46.5 | 3.49 |
| 3 | 3(hcp-cap) | -234.9 | -189.6 | 2.47 |
| 3 | 1(hcp-cap)+1(fcc-cap)+1(top) | -209.7 | -113.9 | 2.72 |
| 3 | 3(fcc-cap) | -209.1 | -112.1 | 2.73 |
| 4 | 1(hcp-cap)+1(fcc-cap)+1(sub-layer)+1(bridge) | -144.2 | +127.8 | 3.39 |
| 4 | 1(peroxo)+1(hcp-cap)+1(top) | -134.9 | +165.0 | 3.51 |
| 4 | 1(peroxo)+1(fcc-cap)+1(top) | -125.8 | +201.5 | 3.63 |
| 4 | 4(top) | -103.7 | +290.1 | 3.92 |
| 5 | 1(sub-layer)+1(peroxo)+1(fcc-cap)+1 (bridge) | -98.2 | +85.9 | 4.17 |
| 5 | 2(peroxo)+1(top) | -68.2 | +233.4 | 4.65 |

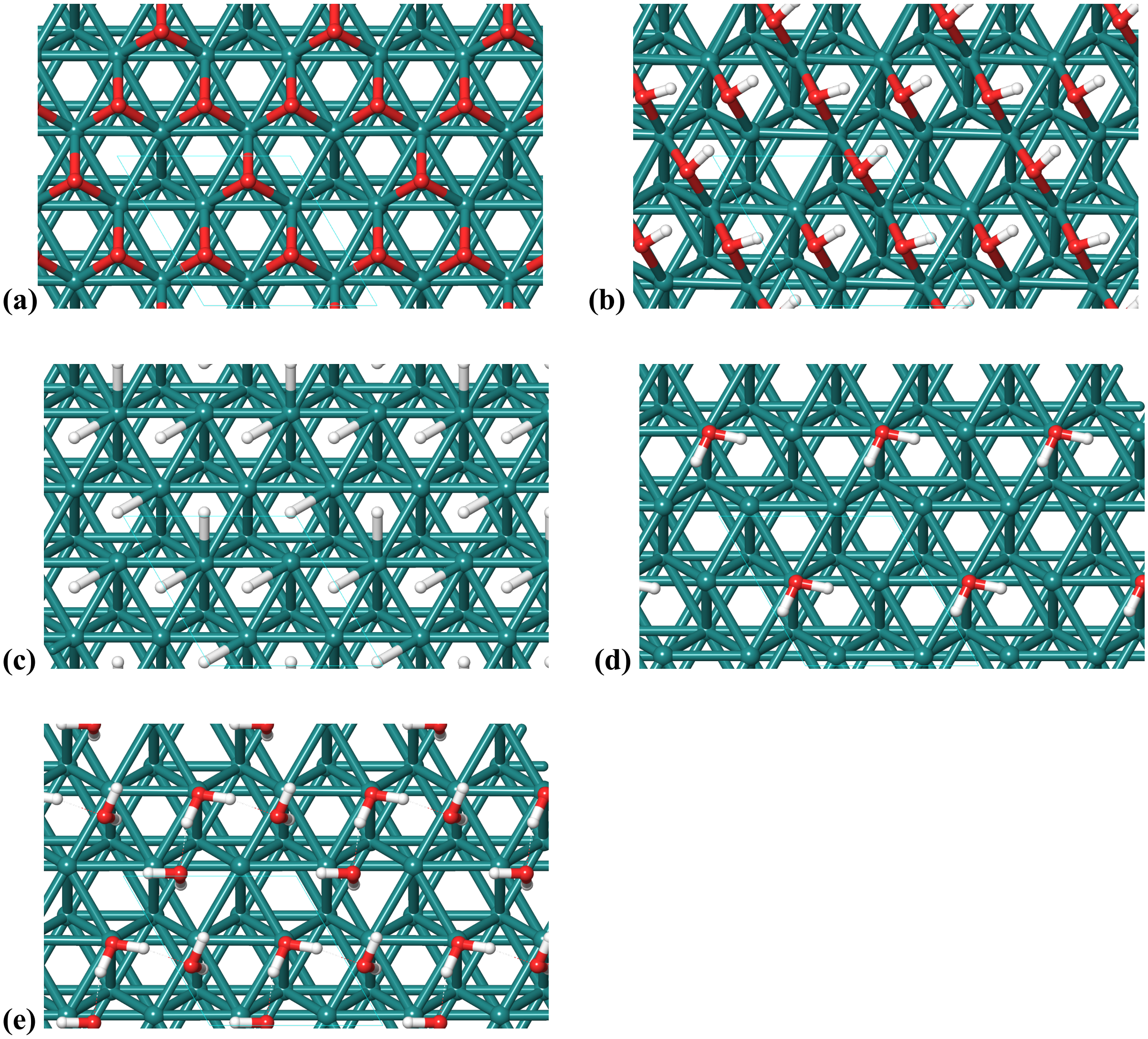

**Figure SI-3.** Top views of periodic expansions of DFT-optimized slab structures of saturating surfaces and intermediates: (a) [3O]*, (b) [3OH]*, (c) [4H]*, (d) [$H_2O$]*, (e) [3$H_2O$]*. Ru=turquoise, O=red, H=white. Figure 2 in the main text shows top and side views of a single cell.

**Section SI-3: Crossover temperature from rigid approximation**

**Table SI-4.** Gibbs reaction free energies within the rigid approximation ($\Delta_r G^{\mathrm{rigid}}$, section 3.4 of main text) over the temperature range of interest and their linear fitting data. Within this approximation $ZPE$ is zero. Linear fitting data and zero point energies ($\Delta_r ZPE$) of the Gibbs reaction free energies from full phonon calculations ($\Delta_r G$).

| **Reaction** | | $\Delta_r \boldsymbol{G}^{\mathbf{rigid}}$ **at** $\boldsymbol{T}$**, kJ/mol** | | | | $\Delta_r \boldsymbol{G}$**, kJ/mol** | |
|---|---|---|---|---|---|---|---|
| | | 0°C | 150°C | 300°C | Slope, $kJ.mol^{-1}.K^{-1}$ | Slope, $kJ.mol^{-1}.K^{-1}$ | $\Delta_r ZPE$, kJ/mol |
| **R1** | $[3O]^*+3/2H_{2(g)} \rightarrow [3OH]^*$ | -7.4 | 0.1 | 8.3 | 0.052 | 0.040 | 10.396 |
| **R2** | $[3O]^*+3H_{2(g)} \rightarrow [3H_2O]^*$ | -32.6 | -17.5 | -1.1 | 0.105 | 0.075 | 20.467 |
| **R3** | $[3O]^*+3H_{2(g)} \rightarrow [H_2O]^*+2H_2O_{(g)}$ | -20.5 | -19.8 | -18.9 | 0.005 | 0.000 | 15.820 |
| **R4** | $[3O]^*+3H_{2(g)} \rightarrow {}^* +3H_2O_{(g)}$ | -16.9 | -23.6 | -30.4 | -0.045 | -0.038 | 14.223 |
| **R5** | $[3O]^*+5H_{2(g)} \rightarrow [4H]^*+3H_2O_{(g)}$ | -64.1 | -60.6 | -56.5 | 0.025 | 0.030 | 18.891 |

**Section SI-4: Structure information**

System (see Table 1 in the main text for labels): *

ATOMIC_POSITIONS crystal

```
Ru 0.33332193 0.66667876 0.51153846
Ru 0.33332154 0.16667847 0.51153846
Ru 0.16668632 0.33331428 0.05009000 0 0 0
Ru 0.33330956 0.16669048 0.14019600 0 0 0
Ru 0.16667177 0.33332880 0.23385499 0 0 0
Ru 0.33332873 0.16667143 0.32774703 0 0 0
Ru 0.16669104 0.83330982 0.42139905 0 0 0
Ru 0.16669065 0.33330953 0.42139905 0 0 0
Ru 0.16668672 0.83331457 0.05009000 0 0 0
Ru 0.33330996 0.66669077 0.14019600 0 0 0
```

```
Ru    0.16667217    0.83332909    0.23385499    0   0   0
Ru    0.33332913    0.66667172    0.32774703    0   0   0
Ru    0.66669073    0.33330953    0.42139905    0   0   0
Ru    0.83332162    0.16667847    0.51153846
Ru    0.66668641    0.33331428    0.05009000    0   0   0
Ru    0.83330965    0.16669048    0.14019600    0   0   0
Ru    0.66667186    0.33332880    0.23385499    0   0   0
Ru    0.83332882    0.16667143    0.32774703    0   0   0
Ru    0.66669094    0.83330982    0.42139905    0   0   0
Ru    0.83332183    0.66667876    0.51153846
Ru    0.66668662    0.83331457    0.05009000    0   0   0
Ru    0.83330986    0.66669077    0.14019600    0   0   0
Ru    0.66667207    0.83332909    0.23385499    0   0   0
Ru    0.83332903    0.66667172    0.32774703    0   0   0
CELL_PARAMETERS bohr
 10.21002600   0.00000000   0.00000000
 -5.11158200   8.83834600   0.00000000
 -0.00000100   0.00000100  43.09096100
```

System: [3O]*

```
ATOMIC_POSITIONS crystal
Ru    0.17239300    0.33516200    0.42240700    0   0   0
Ru    0.33866400    0.16389700    0.51702100
Ru    0.17391700    0.33470500    0.05143000    0   0   0
Ru    0.33995100    0.16786400    0.14193700    0   0   0
Ru    0.17217000    0.33403500    0.23528600    0   0   0
Ru    0.33997000    0.16750100    0.32833200    0   0   0
O     0.17725600    0.33744400    0.57123000
Ru    0.17433900    0.83598700    0.42411700    0   0   0
Ru    0.33860200    0.67250700    0.51704200
```

```
Ru    0.17331600    0.83451900    0.05198700    0   0   0
Ru    0.33994700    0.66780400    0.14192000    0   0   0
Ru    0.17357900    0.83490900    0.23615000    0   0   0
Ru    0.34001200    0.66921100    0.32830000    0   0   0
Ru    0.67562800    0.33521200    0.42242400    0   0   0
Ru    0.84136600    0.16951900    0.51862100
Ru    0.67293800    0.33471500    0.05143300    0   0   0
Ru    0.83996900    0.16787400    0.14231500    0   0   0
Ru    0.67410800    0.33401100    0.23528500    0   0   0
Ru    0.84051400    0.16855800    0.33048500    0   0   0
Ru    0.67527200    0.83772700    0.42245300    0   0   0
Ru    0.84707600    0.67238000    0.51704600
Ru    0.67299200    0.83382000    0.05143300    0   0   0
Ru    0.83990200    0.66780200    0.14191300    0   0   0
Ru    0.67416200    0.83609400    0.23523400    0   0   0
Ru    0.84154900    0.66917800    0.32828800    0   0   0
O     0.67387600    0.33751800    0.57129200
O     0.67380500    0.83390800    0.57128500
CELL_PARAMETERS bohr
 10.21002600   0.00000000   0.00000000
 -5.11158200   8.83834600   0.00000000
 -0.00000100   0.00000100  43.09096100
```

System: [4H]*

```
ATOMIC_POSITIONS crystal
Ru    0.14717600    0.35670900    0.42189200    0   0   0
Ru    0.31386700    0.19014200    0.51439400
Ru    0.14675900    0.35574100    0.05067100    0   0   0
Ru    0.31357200    0.18948600    0.14111300    0   0   0
Ru    0.14732300    0.35645400    0.23464000    0   0   0
```

```
Ru    0.31377900    0.18995400    0.32874600    0   0   0
H     0.48042400    0.02339600    0.56045000
Ru    0.14721100    0.85658700    0.42192800    0   0   0
Ru    0.31381800    0.69006700    0.51442400
Ru    0.14682000    0.85579100    0.05072400    0   0   0
Ru    0.31346900    0.68938200    0.14108100    0   0   0
Ru    0.14726500    0.85650000    0.23465000    0   0   0
Ru    0.31389300    0.68997300    0.32873400    0   0   0
H     0.48053600    0.52350000    0.56046100
Ru    0.64731500    0.35670000    0.42189900    0   0   0
Ru    0.81389700    0.19014900    0.51441700
Ru    0.64681500    0.35583900    0.05067600    0   0   0
Ru    0.81361700    0.18939600    0.14109400    0   0   0
Ru    0.64719200    0.35620400    0.23464700    0   0   0
Ru    0.81404400    0.18979700    0.32873400    0   0   0
H     0.98046300    0.02350500    0.56045300
Ru    0.64721100    0.85667700    0.42197600    0   0   0
Ru    0.81390700    0.69008800    0.51442300
Ru    0.64683500    0.85587600    0.05070600    0   0   0
Ru    0.81356200    0.68930700    0.14113000    0   0   0
Ru    0.64717100    0.85620900    0.23462800    0   0   0
Ru    0.81405200    0.68993100    0.32874100    0   0   0
H     0.98043600    0.52350300    0.56048600
CELL_PARAMETERS bohr
 10.21002600   0.00000000   0.00000000
 -5.11158200   8.83834600   0.00000000
 -0.00000100   0.00000100  43.09096100
```

System: [3OH]*

ATOMIC_POSITIONS crystal

```
Ru     0.11796900    0.30623000    0.42351300    0   0   0
Ru     0.29991100    0.15948600    0.51158500
Ru     0.12206900    0.30924100    0.05174700    0   0   0
Ru     0.28737600    0.14176800    0.14195000    0   0   0
Ru     0.11991000    0.30777200    0.23548900    0   0   0
Ru     0.28658300    0.14069500    0.32902900    0   0   0
O     0.29375400    0.39516100    0.59055000
Ru     0.12784700    0.81146700    0.42420800    0   0   0
Ru     0.29996600    0.63552200    0.51152300
Ru     0.12030100    0.80833200    0.05178900    0   0   0
Ru     0.28748900    0.64192200    0.14193800    0   0   0
Ru     0.11960500    0.80780900    0.23558400    0   0   0
Ru     0.28645200    0.64120800    0.32895900    0   0   0
Ru     0.62993800    0.31393300    0.42258200    0   0   0
Ru     0.78950900    0.14246900    0.51683000
Ru     0.62230000    0.30945200    0.05136100    0   0   0
Ru     0.78769900    0.14208500    0.14209500    0   0   0
Ru     0.62128100    0.30865400    0.23516500    0   0   0
Ru     0.78571500    0.14036900    0.32961400    0   0   0
Ru     0.63005500    0.81059500    0.42254300    0   0   0
Ru     0.81676600    0.65597500    0.51654300
Ru     0.62225900    0.80920700    0.05137700    0   0   0
Ru     0.78696900    0.64156800    0.14180300    0   0   0
Ru     0.62145100    0.80854300    0.23514100    0   0   0
Ru     0.78889500    0.64206900    0.32919700    0   0   0
O     0.79982800    0.39730700    0.58790200
O     0.79956700    0.89842200    0.58790500
H     0.98473700    1.02823000    0.60654000
H     0.98513100    0.45263900    0.60645800
H     0.47929500    0.48848100    0.61026000
CELL_PARAMETERS bohr
```

10.21002600   0.00000000   0.00000000
-5.11158200   8.83834600   0.00000000
-0.00000100   0.00000100  43.09096100

System: [$H_2O$]*

ATOMIC_POSITIONS crystal

| | | | | | | |
|---|---|---|---|---|---|---|
| Ru | 0.04593650 | 0.23350477 | 0.42479416 | 0 | 0 | 0 |
| Ru | 0.21273680 | 0.06647601 | 0.51376903 | | | |
| Ru | 0.04536201 | 0.23274514 | 0.05275446 | 0 | 0 | 0 |
| Ru | 0.21159792 | 0.06531699 | 0.14290238 | 0 | 0 | 0 |
| Ru | 0.04528892 | 0.23254885 | 0.23649923 | 0 | 0 | 0 |
| Ru | 0.21083478 | 0.06527549 | 0.33042495 | 0 | 0 | 0 |
| O | 0.21391017 | 0.54972300 | 0.61802617 | | | |
| Ru | 0.04592564 | 0.73351917 | 0.42442425 | 0 | 0 | 0 |
| Ru | 0.21221968 | 0.56807411 | 0.51590495 | | | |
| Ru | 0.04500331 | 0.73165329 | 0.05273600 | 0 | 0 | 0 |
| Ru | 0.21164760 | 0.56556509 | 0.14269184 | 0 | 0 | 0 |
| Ru | 0.04523261 | 0.73201589 | 0.23651993 | 0 | 0 | 0 |
| Ru | 0.21203132 | 0.56594629 | 0.33054862 | 0 | 0 | 0 |
| Ru | 0.54598101 | 0.23366448 | 0.42314844 | 0 | 0 | 0 |
| Ru | 0.71200775 | 0.06663039 | 0.51377003 | | | |
| Ru | 0.54474322 | 0.23180032 | 0.05261456 | 0 | 0 | 0 |
| Ru | 0.71165425 | 0.06545191 | 0.14296597 | 0 | 0 | 0 |
| Ru | 0.54503552 | 0.23209429 | 0.23666404 | 0 | 0 | 0 |
| Ru | 0.71264808 | 0.06559108 | 0.33043433 | 0 | 0 | 0 |
| Ru | 0.54630150 | 0.73379285 | 0.42440544 | 0 | 0 | 0 |
| Ru | 0.71303072 | 0.56642841 | 0.51366154 | | | |
| Ru | 0.54450058 | 0.73179206 | 0.05276284 | 0 | 0 | 0 |
| Ru | 0.71137700 | 0.56521850 | 0.14297268 | 0 | 0 | 0 |
| Ru | 0.54471324 | 0.73210035 | 0.23651989 | 0 | 0 | 0 |

Ru 0.71270811 0.56727170 0.33039557 0 0 0
H 0.37797631 0.52131087 0.62264583
H 0.04683398 0.35919548 0.62290808
CELL_PARAMETERS bohr
10.21002600 0.00000000 0.00000000
-5.11158200 8.83834600 0.00000000
-0.00000100 0.00000100 43.09096100

System: [3$H_2O$]*

ATOMIC_POSITIONS crystal
Ru 0.04497867 0.23592809 0.41682140 0 0 0
Ru 0.20909126 0.06880120 0.50747851
Ru 0.04513328 0.23614494 0.04678474 0 0 0
Ru 0.21192455 0.06959170 0.13732648 0 0 0
Ru 0.04523351 0.23591293 0.23102525 0 0 0
Ru 0.21290420 0.06825964 0.32477967 0 0 0
O 0.09394664 0.39746873 0.64877741
Ru 0.04348385 0.73492668 0.41864789 0 0 0
Ru 0.21458178 0.57129644 0.50732226
Ru 0.04446068 0.73574369 0.04700555 0 0 0
Ru 0.21191743 0.56943206 0.13722237 0 0 0
Ru 0.04499418 0.73589761 0.23070327 0 0 0
Ru 0.21292461 0.57099905 0.32463578 0 0 0
Ru 0.54538886 0.23654835 0.41917217 0 0 0
Ru 0.71153558 0.06597378 0.50773590
Ru 0.54579267 0.23692320 0.04703309 0 0 0
Ru 0.71194072 0.06953140 0.13716847 0 0 0
Ru 0.54525228 0.23622531 0.23074243 0 0 0
Ru 0.70960311 0.06815060 0.32467630 0 0 0
Ru 0.54619140 0.73495191 0.41907963 0 0 0

| | | | | | | |
|---|---|---|---|---|---|---|
| Ru | 0.71063657 | 0.56577501 | 0.51237788 | | | |
| Ru | 0.54578705 | 0.73556580 | 0.04704704 | 0 | 0 | 0 |
| Ru | 0.71182872 | 0.56946541 | 0.13686667 | 0 | 0 | 0 |
| Ru | 0.54536365 | 0.73598292 | 0.23073840 | 0 | 0 | 0 |
| Ru | 0.71146579 | 0.56884470 | 0.32470682 | 0 | 0 | 0 |
| H | 0.25878770 | 0.57939190 | 0.66045595 | | | |
| H | 0.13753361 | 0.36385076 | 0.60832405 | | | |
| O | 0.33071308 | 0.01185120 | 0.64880534 | | | |
| H | 0.29304202 | -0.05974669 | 0.60784766 | | | |
| H | 0.16155406 | 0.02129155 | 0.66020730 | | | |
| O | 0.71352228 | 0.54470617 | 0.61057926 | | | |
| H | 0.87447482 | 0.51778151 | 0.62424932 | | | |
| H | 0.55043089 | 0.35959894 | 0.62460248 | | | |

CELL_PARAMETERS bohr

| | | |
|---|---|---|
| 10.21002600 | 0.00000000 | 0.00000000 |
| -5.11158200 | 8.83834600 | 0.00000000 |
| -0.00000100 | 0.00000100 | 43.09096100 |